\documentclass[aps,prb,reprint,superscriptaddress]{revtex4-2}
\usepackage[T1]{fontenc}
\usepackage[latin9]{inputenc}
\usepackage{color}
\usepackage{amsmath}
\usepackage{amssymb}
\usepackage{graphicx}
\usepackage[english]{babel}
\usepackage{hyperref}
\usepackage{cases}

\newcommand{\mr}{\mathrm}

\newcommand{\pd}{{\phantom{\dagger}}}

\newcommand{\nd}{n^{\pd}_{d}}

\newcommand{\opa}[2]{#1^{\pd}_{#2}}
\newcommand{\opc}[2]{#1^{\dagger}_{#2}}
\newcommand{\som}[2]{\sum\limits_{#1}^{#2}}
\newcommand{\pro}[2]{\prod\limits_{#1}^{#2}}

\newcommand{\di}[1]{\delta^{}_{#1}}
\newcommand{\za}[2]{\psi^{}_{#1}\left(#2 \right)}
\newcommand{\ra}[2]{\phi^{}_{#1}\left(#2 \right)}

\newcommand\hcancela[2][black]{\setbox0=\hbox{$#2$}\rlap{\raisebox{.42\ht0}{\textcolor{#1}{\hspace{0.15cm}\rule{\wd0}{0.7pt}}}}#2}
\newcommand\hcancelb[2][black]{\setbox0=\hbox{$#2$}\rlap{\raisebox{.42\ht0}{\textcolor{#1}{\hspace{0.24cm}\rule{\wd0}{0.7pt}}}}#2}
\newcommand\hcancelc[2][black]{\setbox0=\hbox{$#2$}\rlap{\raisebox{.42\ht0}{\textcolor{#1}{\hspace{0.09cm}\rule{\wd0}{0.7pt}}}}#2}

\begin{document}

\title{Random matrix theory of charge distribution in disordered quantum impurity models}

\author{Maxime Debertolis}
\affiliation{Institute of Physics, University of Bonn, Nussallee 12, 53115 Bonn, Germany}
\author{Serge Florens}
\affiliation{Univ. Grenoble Alpes, CNRS, Grenoble INP, Institut N\'eel, 38000 Grenoble, France}

\begin{abstract}
We introduce a bare-bone random matrix quantum impurity model, by hybridizing a localized
spinless electronic level with a bath of random fermions in the Gaussian Orthogonal Ensemble (GOE).
While stripped out of correlations effects, this model reproduces some salient features of the
impurity charge distribution obtained in previous works on interacting disordered impurity models.
Computing by numerical sampling the impurity charge distribution in our model, we find a crossover from 
a Gaussian distribution (centered on half a charge unit) at large hybridization, to a bimodal distribution 
(centered both on zero and full occupations of the charge) at small hybridization. In the bimodal
regime, a universal $(-3/2)$ power-law is also observed. All these findings are very well accounted for 
by an analytic surmise computed with a single random electron level in the bath. We also derive an exact functional
integral for the general probability distribution function of eigenvalues and eigenstates, that formally 
captures the statistical behavior of our model for any number $N$ of fermionic orbitals in the bath. 
In the Gaussian regime and in the limit $N\to\infty$, we are able to solve exactly the random matrix theory (RMT) 
for the charge distribution, obtaining perfect agreement with the numerics. Our results could be tested experimentally 
in mesoscopic devices, for instance by coupling a small quantum dot to a chaotic electronic reservoir, and using 
a quantum point contact as local charge sensor for the quantum dot occupation.
\end{abstract}

\maketitle

\section{Introduction}
Disorder plays a central role in condensed matter physics, due to the obvious 
randomness induced by electronic doping in real materials. The conjunction of
disorder with interaction effects opens however many challenges, both conceptually 
and for numerical modeling. For this reason, the study of disordered quantum
impurities offers a simpler vista into such questions. Quantum impurities are 
typically realized by localized electronic levels, traditionally associated to Kondo 
magnetic impurity~\cite{Kondo_Original} that are diluted in a host material~\cite{hewson_1993},
but they can be engineered using various nanoscale platforms, for instance
using semiconducting quantum dots~\cite{KondoGoldhaber,KondoKouwenhoven}.
These quantum impurities display an internal dynamics, complicated by the potentially strong internal 
Coulomb interaction, and acquire at first glance a many-body character due to their hybridization 
to the metallic states in the host material~\cite{WilsonRMP,BullaRMP}. 
In fact, these minimalistic many-body problem turn out to hide a few-body character once expressed 
in a dressed one-particle basis~\cite{he_Lu_NO,Maxime_FewBody,Debertolis_2022}, which led recently 
to the design of ultra-efficient and numerically exact quantum impurity solvers~\cite{Yuriel}. 
In the case of  single quantum impurities in disordered hosts, these natural-orbital methods have opened 
the way to simulate accurately large numbers of random samples~\cite{Debertolis_2022}.
In previous studies~\cite{dobrosavljevic_kondo_1992,
KondoFluctuationsZarand_1996,cornaglia_universal_2006,martin_size_1997,KondoMesoKaul_2003,Lewenkopf_2005,
Kettemann_Mucciolo_PRL_2007,Kettemann_Mucciolo_PRB_2007,zhuravlev_nonperturbative_2007,KondoMesoBurdin_2010,
TransportDirtyKondo_2012,Liu_Burdin_Baranger_Ullmo_EPL_2012,KondoGraphene_2014,Bergmann2015,Slevin_Kettemann_Ohtsuki_2019},
calculations were mostly relying on mean-field approximations and scaling arguments, or
on more expensive numerical simulations like Quantum Monte Carlo, but in any case, very rich physics was uncovered, 
especially in the
statistics~\cite{Liu_Burdin_Baranger_Ullmo_PRB_2012}. These random single-impurity models 
can also be the backbone for disordered Kondo alloys~\cite{KettemannRaikh}, vacancies 
in Moir\'e materials~\cite{KondoVacancyGraphene} or quasicrystals~\cite{KondoQuasiCrystalAndrade,KondoFractal}, 
leading potentially to non Fermi liquid physics~\cite{Miranda_NFL,CastroNetoJones_NFL,KettemannAnnPhys}.
Typically, distribution functions of various observables show multi-modal character
due to the conflict between Kondo screening and Anderson localization.
Clearly, solving the underlying quantum impurity problem (by either controlled or 
approximate methods), and then performing disorder averaging leads little room for 
mathematical understanding of these problems.

In this work, we propose an analytical approach to random quantum impurity problems,
by attacking the simplest model of a quantum impurity in a disordered host. The first 
simplification consists in considering a spinless localized fermionic impurity, which 
can be experimentally realized under a strong in-plane magnetic field in quantum dot devices.
This single-level quantum dot is tunnel coupled by an hybridization $V$ to a disordered 
environment, see Refs.~\cite{Beenakker_RMP,UllmoReview} for general reviews on the statistical 
properties of chaotic electronic devices. Electron-electron interaction effects in the leads will 
also be neglected altogether.
A second crucial simplification into the problem amounts to model the host as 
a random matrix fermionic Hamiltonian under the Gaussian Orthogonal Ensemble (GOE)
with $N$ electronic orbitals~\cite{Meh2004}, instead of taking the tight-binding form with 
local Anderson disorder~\cite{CentralResonantLevel}.
Our random matrix quantum impurity model is easily simulated by solving the one-body impurity problem 
for each disorder realization, and then averaging the statistics. 

We will mostly consider here the probability distribution function $P(n_d)$ of the local charge 
$n_d\in [0,1]$ on the impurity quantum dot, which is a very sensitive measure of its environment. 
Our numerical computations show that $P(n_d)$ displays very rich behavior. For hybridizations $V$ 
that are comparable or larger to the electronic half-bandwidth $2\sqrt{N}\sigma$, the distribution 
takes an unspectacular Gaussian form:
\begin{equation}
\label{eq:Gaussian}
P(n_d) = A \exp\left[-\frac{N (n_d-\frac{1}{2})^2}{2(\sigma_{n_d})^2}\right],
\end{equation}
with $A$ such that the distribution is normalized, and the width $\sigma_{n_{d}}$ is found
to decrease in a non-trivial manner with $V/(2\sqrt{N}\sigma$, and barely depends on $N$. However, for 
hybridizations $V$ that are much smaller than the half-bandwidth $2\sqrt{N}\sigma$, 
the distribution $P(n_d)$ surprisingly turns out 
to be bimodal, with tails diverging as $n_d^{-3/2}$ for $n_d$ close to 0, and as $(1-n_d)^{-3/2}$ for $n_d$ close to 1.
This type of non-Gaussian distribution was reported previously by us, based on advanced numerical calculations 
performed on a model with more realistic Anderson disorder, and including Coulomb 
interaction effects as well. The main goal of the present paper is to provide a general framework to attack 
random matrix quantum impurity models, allowing us to rationalize by analytical means all these interesting 
numerical observations.

\section{Summary of results}
The paper is organized in several distincts sections where the main results are obtained,
and we summarize them here. After introducing the random quantum impurity model in Sec.~\ref{sec:model},
we perform its numerical solution in Sec.~\ref{sec:numerics}, and obtain the data for the impurity charge 
distribution function $P(n_d)$, which displays the Gaussian form~(\ref{eq:Gaussian}) at strong hybridization.
At small hybridization, a bimodal form with universal scaling exponent $(-3/2)$ is found numerically.
In Sec.~\ref{sec:surmise}, we demonstrate that the whole crossover between both types of distributions can 
be very well fitted by a simple ``surmise'' form:
\begin{equation}
P(\nd) = \frac{V}{\sqrt{8\pi}\sigma} 
\text{exp}\left(-\frac{2V^{2}}{\sigma^{2}}\frac{(\nd-\frac{1}{2})^{2}}{\nd(1-\nd)} \right) 
\left[\nd(1-\nd)\right]^{-\frac{3}{2}}\!\!\!,
\end{equation}
that is obtained by replacing the whole electronic bath by a single random level with
mean energy $\sigma$.
At large $V/\sigma$, the Gaussian function peaked at $n_d=1/2$ dominantes the distribution, but
at small $V/\sigma$, the $(-3/2)$ power-law divergence becomes prominent near $n_d=0$ and $1$. 
This power-law is ultimately cut off by the diverging term in the exponential as
$n_d\to0$ or $n_d\to1$, so that the distribution function is normalizable.
Then, in order to obtain analytical results for a finite number of fermions 
in the bath, we derive in Sec.~\ref{sec:PDF} an exact probability distribution function:
\begin{widetext}
\begin{equation}
\label{eq:final_dist}
P\left[E^{}_{\alpha},z^{}_{\alpha}\right] =
\delta\left(V^{2} - \sum_{\gamma=0}^{N}
E^{}_{\gamma} (E^{}_{\gamma}-\epsilon^{}_{d}) z^{}_{\gamma}\right)
\delta\left(\epsilon^{}_{d} - \sum_{\gamma = 0}^{N} E^{}_{\gamma} z^{}_{\gamma}\right) 
\delta\left(1 - \sum_{\gamma=0}^{N}z^{}_{\gamma} \right)
e^{-\frac{1}{4\sigma^{2}} \sum_{\gamma=0}^{N} E^{2}_{\gamma}}
\!\!\!\!\! \prod_{0\leq\gamma<\delta\leq N}{}\!\!\!\!\!
|E^{}_{\delta}-E^{}_{\gamma}| 
\!\!\! \prod_{0\leq\alpha\leq N} \!\!\frac{1}{\sqrt{z^{}_{\alpha}}},
\end{equation}
\end{widetext}
where $E_\alpha$ are the exact eigenenergies of the random matrix impurity model,
$z_\alpha$ are the squared wave function amplitudes of the impurity level in the eigenstates
of the problem, and $\epsilon_d$ is the (non-random) energy of the impurity level.
While the right-most terms ressemble the usual random matrix theory distribution, we emphasize
that the energy levels $E_\alpha$ combine both the impurity degree of freedom and
the set of $N$ electronic orbitals, as also seen by the fact hat all Greek indices here 
run from 0 to $N$.
In addition, a product of three delta functions impose exact constraints on the exact spectrum
and eigenstates, that are mediated via the impurity.
The probability distribution function of the impurity charge at zero temperature can then be computed 
formally by counting only the occupied energy levels:
\begin{equation}
P(\nd) = \!\!\int\limits^{+\infty}_{-\infty}\!\!\mathcal{D}E^{}_{\alpha} 
\!\!\!\int\limits^{+\infty}_{0}\!\!\mathcal{D}z^{}_{\alpha} 
P[E^{}_{\alpha},z^{}_{\alpha}]
\delta\left(\nd - \som{\gamma=0}{N}\Theta(-E^{}_{\alpha})z^{}_{\alpha}\right)\!,
\label{Pnd}
\end{equation}
where $\mathcal{D}E^{}_{\alpha} = \prod^{N}_{\alpha=0}\text{d}E^{}_{\alpha}$ is written for
readability, and the zero temperature Fermi function was replaced by a theta function.
In contrast to previous RMT studies, which focused mostly on spectral properties,
the computation of impurity distributions such as Eq.~(\ref{Pnd}) requires to mix
spectral information with eigenstates distributions, a new step that we take in this
paper.
Finally, in Sec.~\ref{sec:largeN}, we solve the large $N$ limit of Eq.~(\ref{Pnd})
in the Gaussian regime, and find an analytical expression for the width $\sigma_{n_{d}}$ 
of $P(n_d)$ that agrees very well with the width extracted from the numerics.
The expression is not very compact, but can be well approximated by $\sigma_{n_{d}}\simeq
\sqrt{2/\pi-1/2} \sqrt{N}\sigma/V$, and is valid for $V<\sqrt{2N}\sigma$. Indeed, 
the large $N$ solution breaks down for $V>\sqrt{2N}\sigma$, and we find that this critical
value of $V$ corresponds to the point where the impurity is no more hybridized with
the band and exits above the band edge located at energy $E=2\sqrt{N}\sigma$.
An appendice describes some of the computations in more details.

\begin{figure*}[ht]
\includegraphics[width=1.0\textwidth]{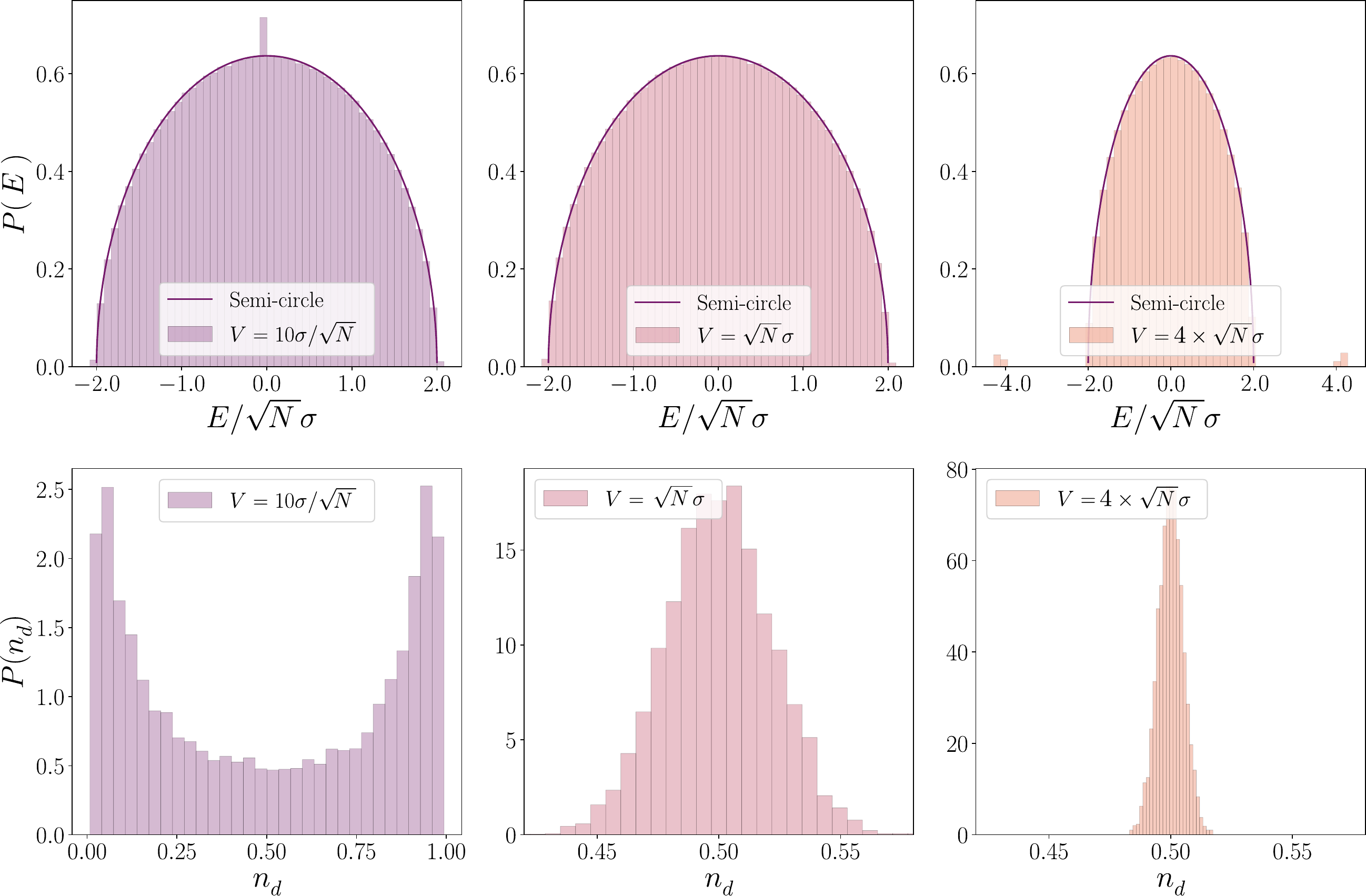}
\caption{
\textbf{Lower panels:} Distribution $P(n_d)$ of the occupation of the impurity level $\nd$ 
obtained for $N=300$ electronic orbitals, $10^4$ GOE disorder realizations. 
From left to right shows three different values of the hybridization $V$, corresponding to 
three different regimes of the problem: weak coupling regime (bimodal charge distribution), 
diluted regime (broad Gaussian distribution), bound states regime (narrow Gaussian distribution), 
see text.
\textbf{Upper panels:} Distribution of energy levels for the corresponding three situations,
confirming the interpretation.
The semi-circle law corresponding to the case of independent bath is plotted for comparison.}
\label{fig:numerics}
\end{figure*}

\section{Random matrix quantum impurity model}
\label{sec:model}
Interacting quantum impurity models can be tackled by a variety of analytical
and numerical techniques. They can also be approximately solved by a non-interacting theory 
with renormalized parameters (slave bosons). For this reason, it is not surprising
that general qualitative features of disordered interacting quantum impurities can be 
present in the case of non interacting fermions. In that case, the spin degree
of freedom can be discarded also. Considering a non-interacting impurity model of spinless fermions will 
thus be our first simplification to the full problem. In addition, disorder in the bath surrounding 
the impurity can be introduced by a large choice of random distribution functions, but the one that 
allows easiest analytical progress is to take a purely random single-particle Hamiltonian, in the 
Gaussian Orthogonal Ensemble. We thus arrive at our random matrix quantum impurity model:
\begin{equation}
\label{H_RMT}
\hat{H}^{}_{\mathrm{imp}} = \epsilon^{}_{d}\opc{d}{}d + V \left(d^{\dagger}c^{}_{1} 
+ \mathrm{H.c.}\right) + \som{i,j=1}{N} \opa{G}{ij} \opc{c}{i}\opa{c}{j},
\end{equation}
with $\epsilon^{}_{d}$ the chemical potential of the impurity level, $V$ the tunneling rate 
of the impurity to a single site of the bath, $N$ the number of electronic orbitals in the bath, 
and $G_{ij}$ a $N\times N$ hermitian matrix with entries taken as independent random variables 
following a centered Gaussian distribution with variance $\sigma$:
\begin{equation}
P\left(G^{}_{ij}\right) =
\frac{1}{\sqrt{2\pi}\sigma}\mathrm{exp}\left(
-\frac{1}{2\sigma^{2}}G^{2}_{ij} \right).
\end{equation}

Our model has hence two competing energy scales, $\sigma$ that controls the strength 
of the disorder, and $V$ which is the tunneling rate between the impurity and the bath. 
In the present context of quantum impurities, this model can be realized by designing a small
single-level quantum dot, associated to the $d^\dagger$ fermions, inside a spin polarized two-dimensional 
electron gas under a large in-plane magnetic field. 
The dot is hybridized via a tunable tunnel barrier into a larger chaotic billiard, whose geometry
is controlled in-situ via electrostatic gates. The charge of the quantum dot can be monitored
via a nearby quantum point contact, acting as a sensitive charge sensor. To the best of our
knowledge, such a simple experiment has not been conducted, but it is clear that maintaining 
the quantum dot half-filled may require fine-tuning from a local plunger gate, due to cross-talk 
effects induced on the impurity when other electrostatic gates are activated.
In a different context, where quantum statistics is not relevant, a similar model was proposed
to describe Mie scattering of light from disordered granular 
media~\cite{Elattari_Kagalovsky_Weidenmuller_1996}. 

In a mean-field treatment of an interacting quantum impurity model, the hybridization $V$ would be 
a renormalized parameter, and would take sample dependent values, due to self-consistency. 
Such effects have been previously considered in Ref.~\cite{Liu_Burdin_Baranger_Ullmo_PRB_2012}, 
but we will keep here $V$ as a fixed number, in order to ease the analytical computations. 
Note also that the impurity level is totally deterministic, namely $\epsilon_d$ is not a random variable.
This is important, otherwise the limit $V/\sigma \ll 1$ would be trivially dominated by
local disorder broadening, and would lead to a Gaussian distribution of the impurity charge. 

In absence of the impurity (corresponding to the strict $V=0$ limit), it is known that the
eigenvalues $E$ of the matrix $G_{ij}$ are distributed for $N$ large enough
according to Wigner's semi-circle law~\cite{Meh2004}:
\begin{equation}
W(E) =
\frac{1}{2\pi\sigma^{2}} \sqrt{4N\sigma^{2}-E^2} \; \Theta(2\sqrt{N}\sigma - |E|),
\end{equation}
which is normalized to the number of states $N$.
Hamiltonian~(\ref{H_RMT}) was previously studied~\cite{Elattari_Kagalovsky_Weidenmuller_1996} 
using standard RMT techniques~\cite{Efetov_1983,Verbaarschot_Weidenmuller_Zirnbauer_1985,Verbaarschot_Zirnbauer_1984},
and it was found that the energy level distribution of the coupled system was only affected at order 
$1/N$ by the presence of the impurity.
However, the charge distribution $P(n_d)$ is expected to remain finite for all $N$ values, and it
will be now investigated from the full numerical solution of the random Hamiltonian~(\ref{H_RMT}).

\section{Charge distribution from numerical sampling}
\label{sec:numerics}
The charge distribution $P(n_d)$ computed by solving model~(\ref{H_RMT}) and averaging over
many disorder realizations is shown in the lower panels of Fig.~\ref{fig:numerics}, together with 
the corresponding eigenenergy distritributions $P(E)$ (upper panels).
These calculations are performed by aligning the impurity level to the Fermi level, namely $\epsilon_d=0$, 
and at zero temperature. Hence, for a given realization, the impurity charge is computed from:
\begin{equation}
\label{nd_rmt}
\nd = \langle \opc{d}{}d \rangle = \som{E^{}_{\alpha}<0}{} z^{}_{\alpha}
= \som{E^{}_{\alpha}<0}{}|\psi^{}_{d}(\alpha)|^{2},
\end{equation}
where $z^{}_{\alpha}$ is the squared amplitude of the exact eigenstate $\psi^{}_d(\alpha)$ on the impurity level 
associated to the occupied energy level $E^{}_{\alpha}$.
In contrast to previous RMT studies of impurity models, $P(n_d)$ requires knowledge of both energies
and wave functions.
In the numerical simulations, we used $N=300$ electronic states in the bath, and the electronic 
half-bandwidth is $2\sqrt{N}\sigma$.
Three different regimes are identified as a function of hybridization $V$ from our data.
\begin{figure}[h!]
\centering
\includegraphics[width=0.45\textwidth]{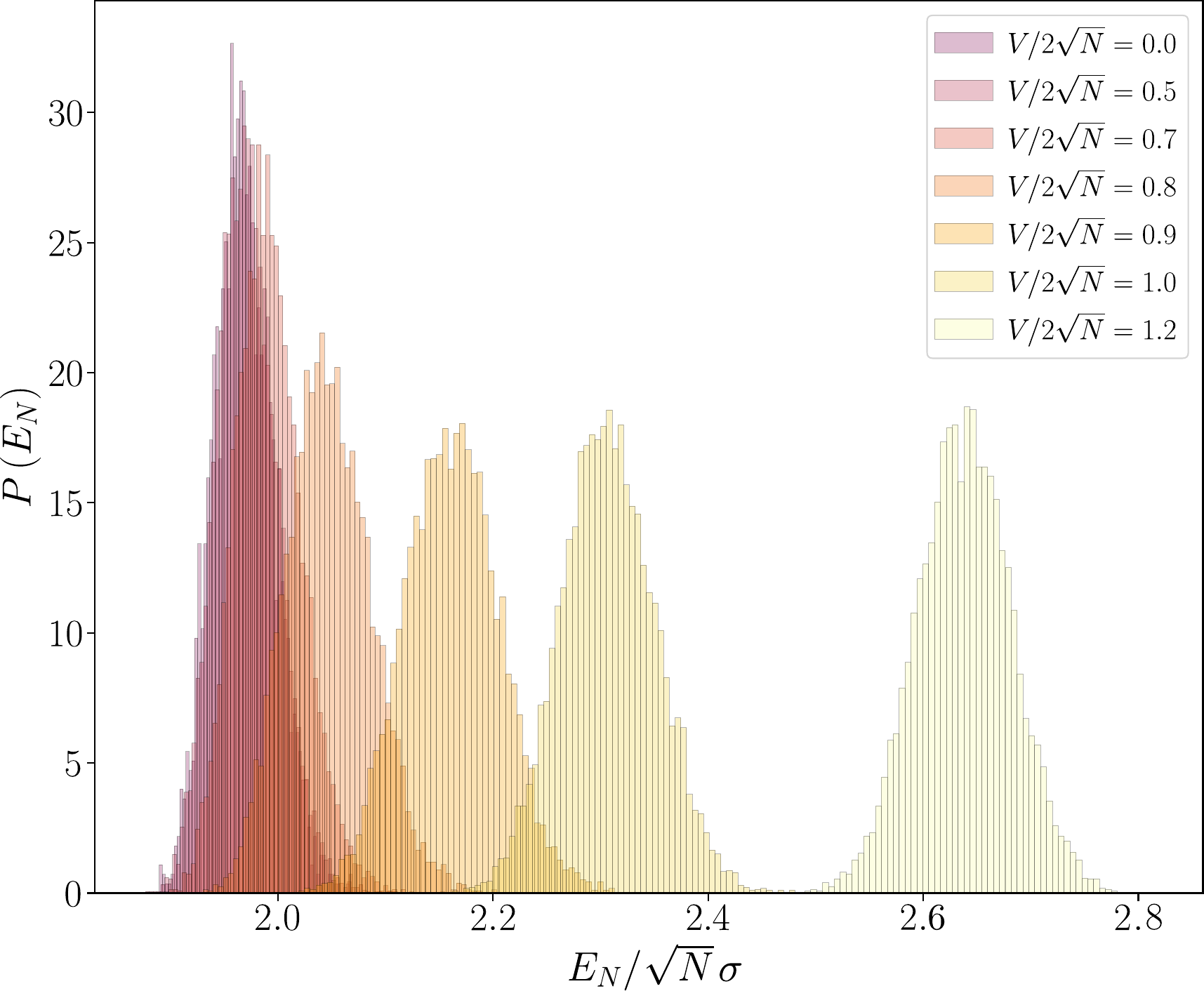}
\caption{Distribution (obtained with $10^4$ realizations) of the maximum energy state $E_N$ = Max$(E_\alpha)$.
For $V<\sqrt{2N}\sigma$, the highest energy state $E_N$ is stuck to the
band edge $2\sqrt{N}\sigma$, showing that the impurity is hybridized within
the band. For $V>\sqrt{2N}\sigma$, the highest energy state exits the band, due
to the formation of a local bound state between the impurity and the $c^\dagger_1$
fermion on site $j=1$.}
\label{fig:BS}
\end{figure}

{\it (i) Weak coupling regime:} 
for $V \propto \sigma/\sqrt{N} = \delta E$, with $\delta E$ the typical energy level spacing,
the impurity is very slightly coupled to its environment, and the square amplitude of the wave function 
on the impurity is close to one, namely $z^{}_{\alpha}\simeq\delta_{\alpha,0}$. As a result, an extra
impurity state near zero energy clearly emerges in the energy distribution on top of the semi-circle, 
see the upper left panel in Fig.~\ref{fig:numerics}.
In this extreme regime, the impurity is easily polarized by its environment, so that $\nd$ is most
likely found close to either $0$ or $1$, depending on the sign of the energy level $E_0$ that is the closest 
to the Fermi level, see the lower left panel in Fig.~\ref{fig:numerics}.

{\it (ii) Diluted regime:} 
for $V < \sqrt{2N}\sigma$, the impurity remains in the band and is well diluted in
the bath. The impurity is coupled to a macroscopic fraction of bath states, and the square amplitudes $z^{}_{\alpha}$ 
become of order $1/N$.
For this reason, the impurity level is diluted over $N$ states and becomes invisible in the
energy distribution at large $N$, which remains very close to the semi-circle law (see the upper middle panel 
in Fig.~\ref{fig:numerics}).
The distribution $P(\nd)$ is now simply centered around $n_d=1/2$ (see the lower middle panel), since one has 
now reached the strong coupling 
regime, and the electrons can efficiently tunnel back and forth from the impurity to the bath.

{\it (iii) Bound state regime:} 
for $V>\sqrt{2N}\sigma$, an impurity bound state emerges out of the electronic band, as seen from the two outliers
outside the semi-circle distribution, see the upper right panel in Fig.~\ref{fig:numerics}.
The escape outside the band is better seen by plotting the distribution of the maximum eigenvalue
Max$(E_\alpha)$, see Fig.~\ref{fig:BS}
For large $V\gg\sqrt{2N}\sigma$,
these bound states at energy $E_\pm \simeq \pm V$ have nearly equal square amplitudes 
$z_\pm\simeq 1/2$, and for this trivial reason, $n_d$ becomes closer and closer to $1/2$. 
Due to the larger value of $V/(2\sqrt{N}\sigma)$, fluctuations into the rest of the bath become rare, 
so that the charge distribution $P(n_d)$ has drastically narrowed down compared to the diluted regime.
In the limit $V\to\infty$, one should obtain $P(n_d) = \delta(n_d-1/2)$, as the wavefunction becomes 
equal to the quantum superposition $(d^\dagger|0\rangle - c^\dagger_1|0\rangle)/\sqrt{2}$,
independently of the disorder values in the bath.

\begin{figure}[h!]
\centering
\includegraphics[width=0.45\textwidth]{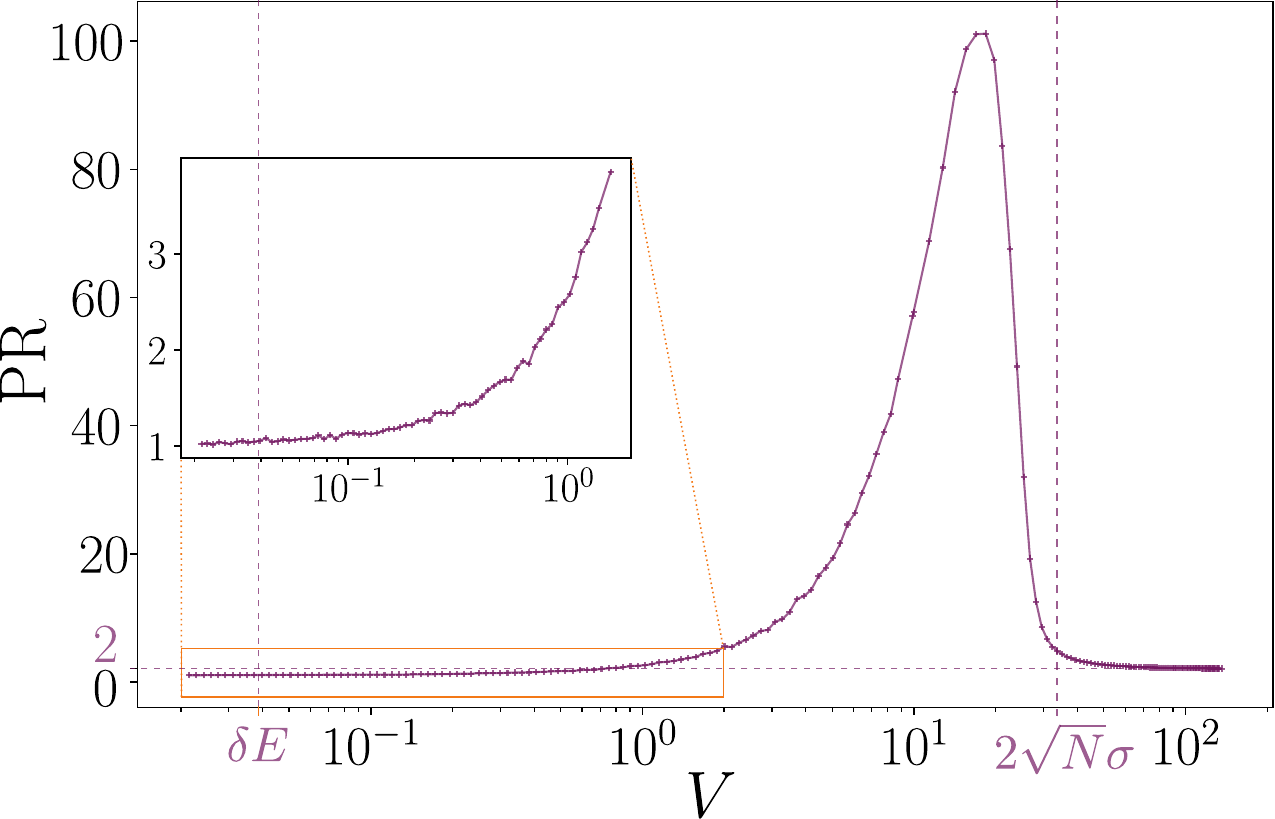}
\caption{Participation ratio (PR) of the impurity wave function as a function of $V$, defined in
Eq.~(\ref{PR}), illustrating
the three regimes of Fig.~\ref{fig:numerics}.
The energy scales $\delta E\propto\sigma/\sqrt{N}$ (level spacing) and $2\sqrt{N}\sigma$ 
(half bandwidth) are shown to delimitate the different crossovers.
The regime of high PR correspond to the case where the impurity is well diluted in the bath.}
\label{fig:PR}
\end{figure}

The three regimes presented above can be characterized quantitatively by the
participation ratio (PR) of the impurity wave function, defined as: 
\begin{equation} 
\label{PR}
\text{PR} =
\overline{ \left(\som{\alpha=1}{N}z^{2}_{\alpha}\right)^{-1}} =
\overline{\left(\som{\alpha=1}{N}|\psi_{d}^{}(\alpha)|^{4}\right)^{-1}}, 
\end{equation} 
which counts the number of eigenstates that effectively hybridize to the impurity, 
averaged over all realizations disorder.
This quantity is reported with respect to $V$ in Fig.~\ref{fig:PR}, each
point being averaged here over $200$ realizations of the random matrix. 
When $V\simeq \sigma/\sqrt{N}=\delta E$, the impurity is almost isolated from
the bath, and the PR equals $1$ since $z^{}_{0}\simeq 1$, as indeed argued above.
When $V$ increases, more and more states become hybridized, and the PR keeps steadily 
increasing to large values of order $N$, confirming that the impurity is diluted,
so that all $z^{}_\alpha\propto 1/N$.
After $V$ has passed the band edge $2\sqrt{N}\sigma$, the PR drastically falls
down to the value $2$, indicating the presence of the two outliers, whose weight is about
$1/2$ each. This confirm the interpretation made above of our numerical data.
We now investigate a simple analytical approach of the random matrix quantum impurity 
model.

\section{Surmise for the bimodal charge distribution}
\label{sec:surmise}

Although the mechanisms behind each of the three regimes found in Fig.~\ref{fig:numerics} are 
driven by different behavior of the hybridization process between the impurity and the bath, 
it turns out that the distribution of the occupation $P(\nd)$ can be obtained by a simple approach. 
Our the numerical results were obtained for fixed and large value $N=300$, in order to reach 
the self-averaging state. Remarkably, we have observed in our simulations that 
the shape of $P(\nd)$ is very little dependent on $N$, even down to $N=1$ (the width of
the charge distribution may however depend on $N$).
We now derive an analytical form of this surmise, by solving the $N=1$ random matrix
impurity model~(\ref{H_RMT}):
\begin{equation}
\hat{H}_{N=1} = V\, (\opc{d}{}\opa{c}{1} + \mathrm{H.c.}) + \epsilon^{}_{1}\opc{c}{1}\opa{c}{1} =
\begin{pmatrix}
0 & V \\
V & \epsilon^{}_{1}
\end{pmatrix},
\label{N1model}
\end{equation}
with $\epsilon^{}_{1}=G_{11}$ a random potential following the normal distribution 
$P(\epsilon^{}_{1}) = (\sqrt{2\pi}\sigma)^{-1}e^{-\epsilon_{1}^{2}/2\sigma^{2}}$. 
We have again fixed $\epsilon_d=0$ for the impurity level. 
The matrix on the right hand side of Eq.~(\ref{N1model}) is readily found
in the Fock basis with one electron in total in the system.

The eigenvalues $\lambda_\pm$ and eigenvectors $\psi_\pm$ of $\hat{H}_{N=1}$ are readily obtained as:
\begin{equation}
\label{eig_toy}
\lambda^{}_{\pm} = \frac{\epsilon_{1}^{}}{2} \pm \frac{1}{2}\sqrt{\epsilon_{1}^{2} + 4V^{2}}, \quad \psi^{}_{\pm} 
= \frac{1}{(V^{2}+\lambda^{2}_{\pm})^{1/2}}
\begin{pmatrix}
V \\
\lambda^{}_{\pm}
\end{pmatrix},
\end{equation}
Using Eq.~\eqref{nd_rmt} for the impurity charge $\nd$, its distribution reads:
\begin{equation}
\label{pnd_1}
P(n^{}_{d}) = \int\limits^{+\infty}_{-\infty} \text{d}\epsilon_{1}
P(\epsilon_{1})\,\delta\left(n^{}_{d}-\sum^{}_{\alpha=\pm}\Theta\big(-\lambda^{}_{\alpha}\big)|\psi_{\alpha}|^2 \right).
\end{equation}
Inserting expressions~(\ref{eig_toy}) and computing the integral in Eq.~\eqref{pnd_1} leads 
to the desired distribution of $\nd$:
\begin{equation}
P(\nd) = \frac{V}{\sqrt{8\pi}\sigma} 
\text{exp}\left(-\frac{2V^{2}}{\sigma^{2}}\frac{(\nd-\frac{1}{2})^{2}}{\nd(1-\nd)} \right) 
\left[\nd(1-\nd)\right]^{-\frac{3}{2}}\!\!\!.
\label{surmise}
\end{equation}

\begin{figure}[h!]
\centering
\includegraphics[width=0.475\textwidth]{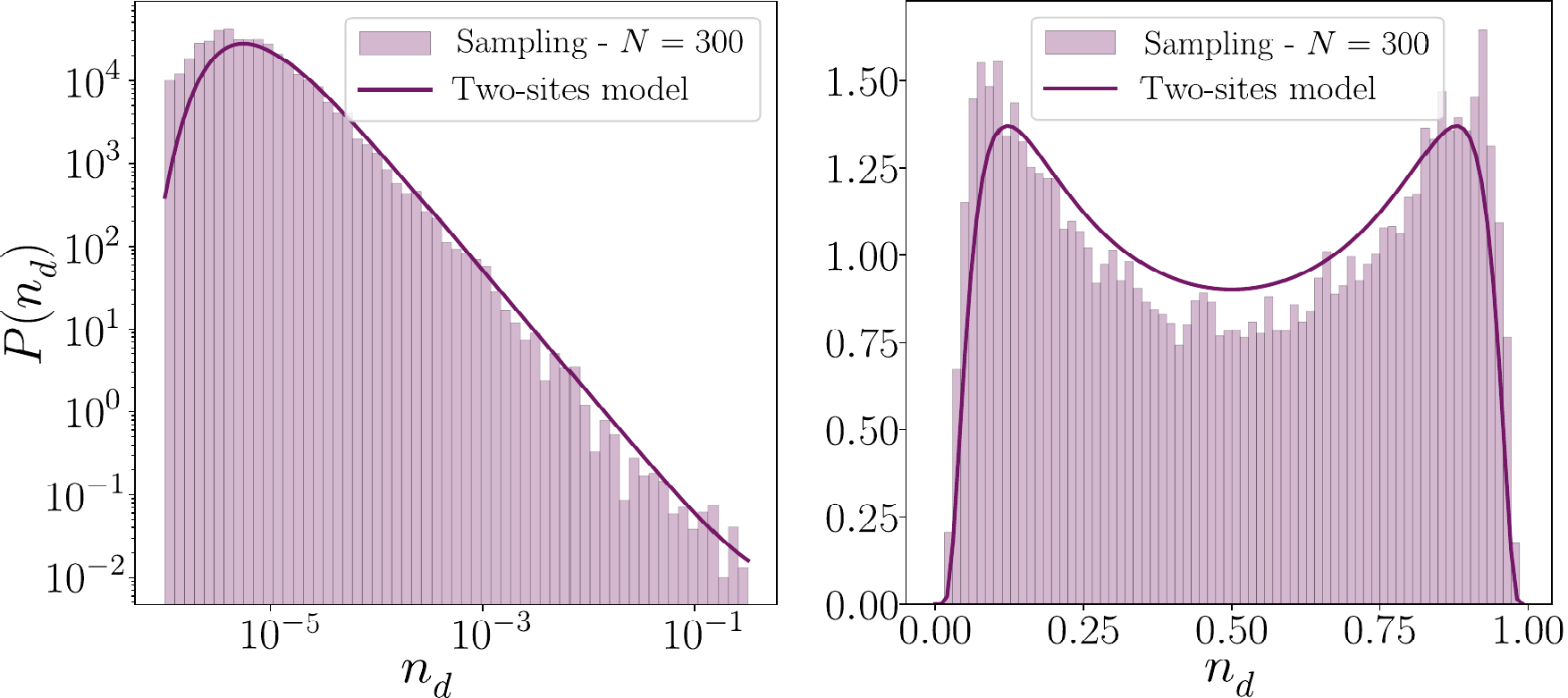}
\caption{Charge distribution $P(n_d)$ of the full random matrix impurity model with $N=300$
electronic orbitals, with the histograms computed by numerical sampling with $10^4$ disorder
realizations, for $V/\sigma\ll1$ ({\bf left panel}, on log scale) and $V/\sigma \lesssim1$ 
({\bf right panel}, on linear scale).
The comparison to the analytical surmise Eq.~(\ref{surmise}), using $V/\sigma$ as a fitting
parameter, is remarkably good, especially the universal $n_d^{-3/2}$ power law is well 
established in the numerical data at $V/\sigma\ll1$.}
\label{fig:comparison}
\end{figure}

This expression shows the competition of two terms.
At strong coupling, namely $V/\sigma\gg1$, the exponential term dominates
and $P(\nd) \sim \text{exp}[-\frac{2V^{2}}{\sigma^{2}}(\nd-\frac{1}{2})^{2}]$ which is a
Gaussian of mean $1/2$ and of deviation $\sigma/2V$. Clearly, it will fit the bell-shaped
charge distributions obtained in the diluted and bound state regimes (bottom center and bottom
right panels of Fig.~\ref{fig:numerics}), upon ajusting the width $\sigma$ of the two site
model, since the actual width does depend on $N$.
At weak coupling, namely $V/\sigma\ll1$, we find $P(\nd) \sim (\nd(1-\nd))^{-\frac{3}{2}}$,
leading to an universal (-3/2) power laws for $n_d$ close to $0$ or $1$. 
The $1/{\nd(1-\nd)}$ factor in 
the exponential term prevents the distribution to ultimately diverge for $\nd\to0$ or $\nd\to1$.
In Fig.~\ref{fig:comparison}, we compare the analytical surmise~(\ref{surmise}) to the charge
distribution obtained for $N=300$ orbitals in the $V/\sigma\ll1$ regime (left panel) and
$V/\sigma\lesssim1$ regime (right panel). The agreement is remarkably good, despite the fact
that the analytical calculation corresponds to the strict $N=1$ limit. In particular,
for $V/\sigma \ll 1$ (left panel), the power law $(\nd(1-\nd))^{-3/2}$ is well observed 
in the numerics, making it a robust universal feature of the full model. Slight quantitative
deviations near the edges are however seen, unsurprisingly. In addition, for $V/\sigma\lesssim1$,
we recover the long tails yielding the previously observed plateau of $P(\nd)$ around $\nd=1/2$. 

The two-site model thus provides a simple analytical formula describing $P(\nd)$ at least
qualitatively in every regime of the full model, and even quantitatively in some situations.
However, some other observables are poorly reproduced, so that a solution for the full problem 
is still needed. 
In the following section, the joint probability distribution of eigenvalues and eigenvectors of 
the full Hamiltonian~\eqref{H_RMT} is exactly built for any value of $N$, as a first step
for a RMT solution of the problem

\section{Exact formula for the probability distribution functions}
\label{sec:PDF}

This section aims to find the exact joint probability distribution of the set of eigenvalues
$E_\alpha$ and eigenvectors $\psi_\alpha$ of the random matrix impurity model. 
The starting point is the distribution of the unperturbed eigenvalues $e_\alpha$ and 
eigenvectors $\phi_\alpha$ of the random hermitian matrix $G_{ij}$ of the bath,
that is known to obey the random matrix Gaussian orthogonal ensemble (GOE)~\cite{Meh2004}:
\begin{eqnarray}
\nonumber
P_\mr{RMT}[e^{}_{\alpha},r^{}_{\alpha}] &=& \!\!\!\! \pro{1\leq\alpha<\beta\leq N}{}
\!\!\!\!\! |e^{}_{\alpha} - e^{}_{\beta}| \;\;\!
e^{-\frac{1}{4\sigma^{2}}\som{\beta=0}{N}e_{\beta}^{2}}
\pro{1\leq\alpha\leq N}{} \frac{1}{\sqrt{r^{}_{\alpha}}} \\
\label{distrib0}
&& \times \delta\left(\som{\beta=1}{N}r^{}_{\beta}-1\right),
\end{eqnarray}
where $r^{}_{\alpha}=|\phi(\alpha)|^{2}$ is the square amplitude, and the delta distribution
imposes the normalizationn of the wave function.
A similar study to ours was performed for a random matrix model with a local 
defect~\cite{Aleiner_Matveev_1998,Bogomolny_2017}. 
The perturbation corresponded there to a deterministic potential acting on a given entry of 
the random matrix, leading to a rank-one perturbation of the random matrix of rank $N$. 
In our random matrix quantum impurity system, the perturbation is of rank two, as it adds
an extra physical orbital into the problem.
The matrix model~\eqref{H_RMT} can be reformulated in a more convenient way using a matrix form
$H^{}_{ij}$, with indices $i,j=0,\ldots,N$, such that:
\begin{equation}
\label{model_rmt}
\begin{cases}
H^{}_{00} = \epsilon^{}_{d},\\
H^{}_{0j} = H^{}_{j0} = V \quad &\forall j\neq 0,\\
H^{}_{ij} = G^{}_{ij} \quad &\text{otherwise}.
\end{cases}
\end{equation}
\vspace{0.0cm}\\
The entries $G^{}_{ij}$ for $i,j>0$ are distributed following a Gaussian distribution of zero mean
and of deviation $\sigma$, and are real and symmetric, in order to work in the GOE.
If we add the impurity site $i=0$ with energy $e_0$ and square amplitude $r_0$, we thus obtain
in the decoupled limit $V=0$ a new distribution:
\begin{equation}
P[e^{}_{\alpha},r^{}_{\alpha}] = P_\mr{RMT}[e^{}_{\alpha},r^{}_{\alpha}] 
\delta\left(e^{}_{0}-\epsilon^{}_{d}\right)\delta\left(r^{}_{0}\right),
\end{equation}
due to the constraints required in Eq.~\eqref{model_rmt}.

We want to compute the charge distribution of the impurity $P(\nd)$, expressed
as $\nd = \sum_{\alpha}|\za{0}{\alpha}|^{2}\Theta(-E_\alpha)$, where $\za{0}{\alpha}$ is the site $0$
component of the eigenvector $\alpha$ of the full matrix $H_{ij}$, and $E^{}_{\alpha}$ is its
associated eigenvalue. We will denote $z^{}_{\alpha} = |\psi^{}_{0}(\alpha)|^{2}$ the square amplitude 
of this eigenvector. In order to obtain the joint probability distribution function of the
exact energies and amplitudes for the coupled system, denoted
$P[E^{}_{\alpha},z^{}_{\alpha}]$, we need to perform a change of variables between the $N^{2}$ 
variables ($e_\alpha, r_\alpha$) of the decoupled problem to the $N)^{2}$ variables 
($E_\alpha, z_\alpha$) of the coupled problem. There are however $(N+1)^{2}$ entries in $H_{ij}$
according to Eq.~(\ref{model_rmt}), but since the first line and the first column are known, 
they are not associated to random variables, and are fixed by the two constraints imposed on 
the eigenvalues and eigenvectors of $H$. The details of this calculation are presented 
in Appendix~\ref{AppA}, and the final expression for the eigenenergy and eigenstate distribution 
function reads:
\begin{widetext}
\begin{equation}
\label{final_dist}
P\left[E^{}_{\alpha},z^{}_{\alpha}\right] =
\delta\left(V^{2} - \sum_{\gamma=0}^{N}
E^{}_{\gamma} (E^{}_{\gamma}-\epsilon^{}_{d}) z^{}_{\gamma}\right)
\delta\left(\epsilon^{}_{d} - \sum_{\gamma = 0}^{N} E^{}_{\gamma} z^{}_{\gamma}\right) 
\delta\left(1 - \sum_{\gamma=0}^{N}z^{}_{\gamma} \right)
e^{-\frac{1}{4\sigma^{2}} \sum_{\gamma=0}^{N} E^{2}_{\gamma}}
\!\!\!\!\! \prod_{0\leq\gamma<\delta\leq N}{}\!\!\!\!\!
|E^{}_{\delta}-E^{}_{\gamma}| 
\!\!\! \prod_{0\leq\alpha\leq N} \!\!\frac{1}{\sqrt{z^{}_{\alpha}}},
\end{equation}
\end{widetext}
This equation is the main result of this section, and is exact (up to a norm) for any value
of the hybridizationparameter $V$, and for all system sizes $N$. 
We have also checked from the numerics performed in Sec.~\ref{sec:numerics}
that, for each disorder realization, the three delta functions 
in Eq.~(\ref{final_dist}) above are exactly obeyed, confirming the
correctness of our analytical result. 
The first constraint, $V^{2} = \sum_{\gamma=0}^{N}E^{}_{\gamma} (E^{}_{\gamma}-\epsilon^{}_{d})
z^{}_{\gamma}$, corresponds actually to the constraint seen in Eq.~\eqref{distrib0}, re-expressed 
with the new variables. The two other constraints are imposed to reduce the number of $(N+1)^{2}$ variables 
in the new basis, to the $N^{2}$ random variables of the original problem.

The impurity charge distribution can be computed formally from:
\begin{equation}
P(\nd) = \!\!\int\limits^{+\infty}_{-\infty}\!\!\mathcal{D}E^{}_{\alpha} 
\!\!\!\int\limits^{+\infty}_{0}\!\!\mathcal{D}z^{}_{\alpha} 
P[E^{}_{\alpha},z^{}_{\alpha}]
\delta\left(\nd - \som{\gamma=0}{N}\Theta(-E^{}_{\alpha})z^{}_{\alpha}\right)\!,
\label{eq:Pnd}
\end{equation}
where $\mathcal{D}E^{}_{\alpha} = \prod^{N}_{\alpha=0}\text{d}E^{}_{\alpha}$ is written for
compactness.
The main difficulty at this point is how to deal correctly with the different non-linear
constraints in Eq.~(\ref{final_dist}). 
In the upcoming and final section, exact analytical calculations of the charge distribution function 
are presented, using the large $N$ limit.

\section{Large N solution}
\label{sec:largeN}

We finally compute analytically the charge distribution~(\ref{eq:Pnd}) in the large $N$ limit,
assuming the potential acting on the impurity to be unbiased, namely $\epsilon_d=0$.
The first step is to Fourier transform the three delta distributions in Eq.~(\ref{final_dist}) 
and the fourth one in Eq.~(\ref{eq:Pnd}), using
$ \delta\left(x\right) = \int^{+\infty i}_{-\infty i} \frac{\text{d}\lambda}{2\pi} \;
\text{exp}\left [ \lambda x\right] $
where $\lambda$ was written as an purely imaginary variable.
Integrating over the $z_\alpha$ variables in Eq.~(\ref{final_dist}) can then be done
exactly, and we obtain a formally exact expression:
\begin{equation}
\label{Pnd1}
P(\nd) = \int\limits^{+\infty}_{-\infty}\mathcal{D}E^{}_{\alpha}
\int\limits_{-\infty i}^{}\hspace{-0.2cm}\int\hspace{-0.1cm}\int\hspace{-0.2cm}\int\limits^{+\infty i}_{}
\frac{\text{d}\lambda\,\text{d}\nu\,\text{d}\rho\,\text{d}\mu}{(2\pi)^{7/2}}
e^{-F[E_\gamma,\lambda,\mu,\nu,\rho]},
\end{equation}
with the ``free energy'' functional:
\begin{eqnarray}
\nonumber
F[E_\gamma,\lambda,\mu,\nu,\rho] &=&
\som{\gamma=0}{N} \frac{E^{2}_{\gamma}}{4\sigma^{2}} \;\;-\!\!\!
\som{0\leq\delta<\gamma\leq N}{} \text{ln}\left|E^{}_{\delta}-E^{}_{\gamma}\right| \\
\nonumber
&&\hspace{-0.3cm} +\frac{1}{2}\som{\gamma=0}{N}\, 
\text{ln}\left|\lambda E^{2}_{\gamma} + \nu E^{}_{\gamma} + \rho + \mu \Theta(-E^{}_{\gamma})
\right|\\
&& - \lambda V^{2} - \rho - \mu\, \nd.
\label{Free1}
\end{eqnarray}

In the limit of large $N$, the Laplace method is used to compute the integrals
over the energies $E_\alpha$. We follow here Ref.~\cite{Brezin_Itzykson_Parisi_Zuber_1978}, 
in which a random matrix model perturbed by a $\phi^{4}$ bulk interaction was solved. 
In our impurity model, the scaling of each terms has to be carefuly accounted for. We know that 
energies of the random matrix scale as $E^{}_{\alpha}\propto \sqrt{N}$, and we want to
consider the diluted regime with $V\propto \sqrt{N}$, see middle panels of Fig.~\ref{fig:numerics}. 
In order for all the terms inside the logarithm in Eq.~(\ref{Free1}) to be of comparable
size, we thus need to scale the Lagrange multipliers as 
$\lambda\propto 1$, $\nu\propto\sqrt{N}$, $\rho \propto N$, and $\mu \propto N$. 
With these scalings, the saddle-point equations $\partial_{E^{}_{\alpha}} P(\nd) = 0$ read for
$E_\alpha>0$:
\begin{equation}
\som{\substack{0\leq\beta\leq N, \\ 
\beta\neq\alpha}}{} \frac{2}{E^{}_{\alpha}-E^{}_{\beta}} = \frac{E^{}_{\alpha}}{2\sigma^{2}} 
+ \frac{1}{2} \frac{2 \lambda E^{}_{\alpha} + \nu}{\lambda E^{2}_{\alpha} + \nu E^{}_{\alpha} + \rho},
\end{equation}
and it is easy to check that the second term in the right hand side, induced by the impurity,
is $1/N$ smaller than the other two terms, and can be neglected when $N$ is large (the same
argument applies for $E_\alpha<0$). Unless the density of state is strongly peaked
near $E_\alpha=0$ (weak coupling regime), we deduce that the density of states is
barely affected by the impurity at large $N$, and this corresponds to the diluted regime 
shown in the top middle panel of Fig.~\ref{fig:numerics}.

Going to the continuous limit $E^{}_{\alpha}\rightarrow E(\alpha/N)$,
and $1/N \sum^{N}_{\alpha} \rightarrow \int^{1}_{0}\text{d}x$, we obtain the 
expression~\cite{Brezin_Itzykson_Parisi_Zuber_1978}:
\begin{equation}
\hcancela{\int}\limits^{1}_{0} \frac{\text{d}y}{E(x)-E(y)} = \frac{1}{4N\sigma^{2}}E(x),
\end{equation}
where $\hcancelc{\int}$ represents the principal part of the integral. At this point, 
the density of states $u(E)=\text{d}y/\text{d}E(y)$ is introduced, normalized on an interval
$[-2a,2a]$ which gives:
\begin{equation}
\label{eqq1}
\hcancelb{\int}\limits^{2a}_{-2a} \text{d}F \, \frac{u(F)}{E-F} = \frac{1}{4N\sigma^{2}}E, \quad \forall |E|\leq 2a.
\end{equation}
It is direct to check that the density of states solving this equation~\cite{Meh2004} is given by a semi-circle 
law in the support $[-2\sqrt{N}\sigma, 2\sqrt{N}\sigma]$, namely:
\begin{equation}
u(E) = \frac{1}{2\pi \sigma^2} \sqrt{4N\sigma^{2} - E^{2}} \; \Theta(2\sqrt{N}\sigma-|E|).
\end{equation}
At saddle-point level, we can remove all the integrals over $E^{}_{\alpha}$ in Eq.~(\ref{Pnd1}):
\begin{equation}
\label{Pnd2}
P(\nd) = \int\limits_{-\infty i}^{}\hspace{-0.2cm}\int\hspace{-0.1cm}\int\hspace{-0.2cm}\int\limits^{+\infty i}_{}
\frac{\text{d}\lambda\,\text{d}\nu\,\text{d}\rho\,\text{d}\mu}{(2\pi)^{7/2}} \; 
e^{-F[\lambda,\mu,\nu,\rho]},
\end{equation}
and express the energy dependence as a single integral over the density of states:
\begin{eqnarray}
\nonumber
F[\lambda,\mu,\nu,\rho] &=& 
\int\limits_{-2\sqrt{N}\sigma}^{0} \hspace{-0.3cm}\text{d}E \, u(E)
\frac{1}{2}\,\text{ln}\left(\lambda E^{2} + \nu E + \rho + \mu \right)\\
\nonumber
&&+\!\!\! \int\limits^{2\sqrt{N}\sigma}_{0} \hspace{-0.3cm}\text{d}E \, u(E) 
\frac{1}{2}\, \text{ln}\left(\lambda E^{2} + \nu E + \rho \right)\\
&& - \lambda V^{2} - \rho - \mu\, \nd.
\end{eqnarray}
From the scaling of the parameters $V,\lambda,\mu,\nu,\rho$, we see that 
this free energy is proportional to $N$, so that at large $N$ the integration 
over the four Lagrange multipliers can be treated 
by steepest descent,
$\partial F[\lambda,\mu,\nu,\rho]/\partial x=0$, with $x=\lambda,\mu,\nu,\rho$,
which gives:
\begin{eqnarray}
\nonumber
V^2 \, &=& \hspace{-0.2cm} \int\limits_{-2\sqrt{N}\sigma}^{0} \hspace{-0.4cm} \text{d}E \, 
\frac{u(E) E^2/2}{\lambda E^{2} + \nu E + \rho + \mu} \, 
+ \hspace{-0.2cm} \int\limits^{2\sqrt{N}\sigma}_{0} \hspace{-0.3cm} \text{d}E \, 
\frac{u(E) E^2/2}{\lambda E^{2} + \nu E + \rho},\\ 
\label{saddles1}
n_d \, &=& \hspace{-0.2cm} \int\limits_{-2\sqrt{N}\sigma}^{0} \hspace{-0.4cm} \text{d}E \, 
\frac{u(E)/2}{\lambda E^{2} + \nu E + \rho + \mu}, \\
\nonumber
0 \, &=& \hspace{-0.2cm} \int\limits_{-2\sqrt{N}\sigma}^{0} \hspace{-0.4cm} \text{d}E \, 
\frac{u(E) E/2}{\lambda E^{2} + \nu E + \rho + \mu} 
+\hspace{-0.2cm} \int\limits^{2\sqrt{N}\sigma}_{0} \hspace{-0.3cm} \text{d}E \, 
\frac{u(E) E/2}{\lambda E^{2} + \nu E + \rho},\\ 
\nonumber
1 \, &=& \hspace{-0.2cm} \int\limits_{-2\sqrt{N}\sigma}^{0} \hspace{-0.4cm} \text{d}E \, 
\frac{u(E)/2}{\lambda E^{2} + \nu E + \rho + \mu}
+\hspace{-0.2cm} \int\limits^{2\sqrt{N}\sigma}_{0} \hspace{-0.3cm} \text{d}E \, 
\frac{u(E)/2}{\lambda E^{2} + \nu E + \rho}.
\end{eqnarray}
After some manipulations, and using the fact that the density of states $u(E)$ is normalized to $N$, 
we find the symmetrized system of equations:
\begin{eqnarray}
\nonumber
\lambda V^{2} &+& \rho +\mu n_d = N/2, \\
\label{saddles2}
n_d \, &=& \hspace{-0.2cm} \int\limits_0^{2\sqrt{N}\sigma} \hspace{-0.3cm} \text{d}E \, 
\frac{u(E)/2}{\lambda E^{2} - \nu E + \rho + \mu}, \\
\nonumber
0 \, &=& 
\hspace{-0.2cm} \int\limits^{2\sqrt{N}\sigma}_{0} \hspace{-0.3cm} \text{d}E \, 
\frac{ (\mu-2\nu E) u(E) E/2}
{(\lambda E^{2} + \nu E + \rho) (\lambda E^{2} - \nu E + \rho + \mu)},\\
\nonumber
2n_d-1 \, &=& \hspace{-0.2cm} \int\limits^{2\sqrt{N}\sigma}_{0} \hspace{-0.3cm} \text{d}E \, 
\frac{(\mu-2\nu E) u(E)/2}
{(\lambda E^{2} + \nu E + \rho) (\lambda E^{2} - \nu E + \rho + \mu)}.
\end{eqnarray}

These non-linear equations do not have explicit analytical solution in general.
However, we can make further progress by analyzing them in the vicinity of $n_d=1/2$,
where a Gaussian peak was found to describe $P(n_d)$ in the diluted regime of the impurity,
see top middle panel in Fig.~\ref{fig:numerics}.
Indeed, due to particle-hole symmetry (in average) for $n_d=1/2$, we see that $\mu=\nu=0$ satisfies
automatically the last two of the saddle point equations in Eqs.~(\ref{saddles2}), and therefore 
we have simply 
$\rho = N/2 -\lambda V^2$, with
$\int^{2\sqrt{N}\sigma}_{0} \text{d}E \, 
\frac{u(E)/2}{\lambda E^{2} + \rho} = 1/2$ to be solved for $\lambda$.
Performing the integration over $E$ exactly, we obtain the equation:
\begin{equation}
\label{simple}
\sqrt{1+4\lambda\sigma^2/(1/2-\lambda V^2/N)} = 1 + 4\lambda\sigma^2,
\end{equation}
recalling that $V\propto N$ and $\lambda\propto 1$, for the saddle point to be valid
at large $N$. Squaring Eq.~(\ref{simple}), we readily find the simple solution
$\lambda = N/(2V^2)- 1/(2\sigma^2)$, that applies for $n_d=1/2$.
Noting that the left hand side of Eq.~(\ref{simple})
must be positive, we obtain the condition that $1 + 4\lambda\sigma^2>0$, which implies
$|V|<\sqrt{2N}\sigma$. This is precisely the condition that the impurity level
remains inside the electronic band, as seen in Fig.~\ref{fig:BS}, and in agreement 
with our initial hypothesis that the impurity is diluted.

\begin{figure}[h!]
\centering
\includegraphics[width=0.5\textwidth]{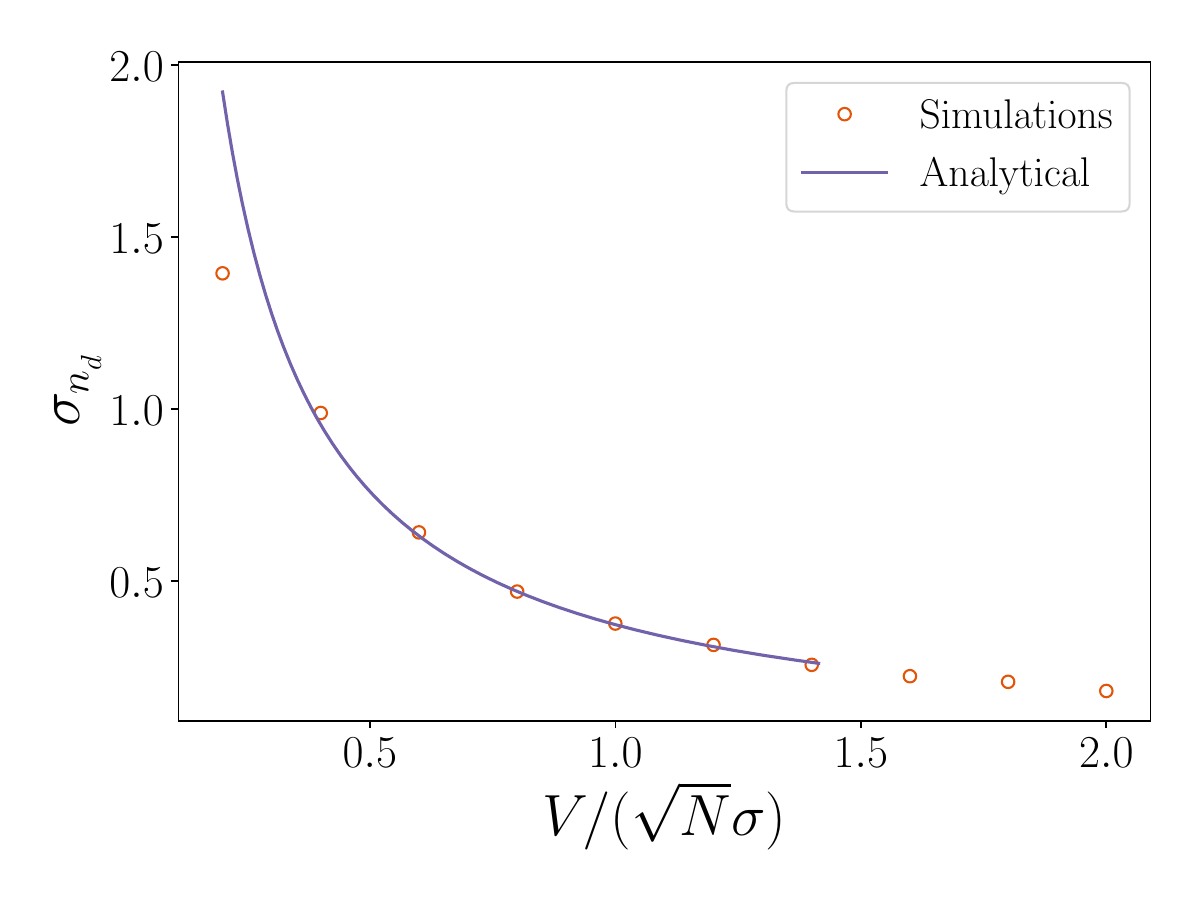}
\caption{Comparison of the large $N$ analytical expression~(\ref{derivative}) for the width $\sigma_{n_{d}}$ 
of $P(n_d)$ (solid line) obtained in the diluted regime $\sigma/\sqrt{N} \ll V < \sqrt{2N}\sigma$, 
to the width extracted from the numerics (dots) using 2000 disorder realizations 
and fitting the histogram of the charge distribution with the Gaussian formula~(\ref{Gauss}).
The large $N$ theory breaks down for
$V/(\sqrt{N}\sigma)>\sqrt{2}$ and $V/(\sqrt{N}\sigma)\ll1$, as expected.}
\label{fig:largeN}
\end{figure}

The final trick to find $P(n_d)$ is to express it in the close vicinity of $n_d=1/2$. At
the saddle point level, we have $P(n_d) = A e^{-F[\lambda,\mu,\nu,\rho]}$, where
the saddle point values satisfying Eq.~(\ref{saddles2}) must be inserted, and $A$ is
a normalization factor. Due to the fact that
$\partial F[\lambda,\mu,\nu,\rho]/\partial x=0$, with $x=\lambda,\mu,\nu,\rho$,
and using $\mu=0$ for $n_d=1/2$, a second order Taylor expansion gives a Gaussian
charge distribution:
\begin{equation}
P(n_d) = P(n_d=1/2) \exp\left[-\frac{N(n_d-1/2)^2}{2(\sigma_{n_{d}})^2}\right],
\label{Gauss}
\end{equation}
with the width expressed as
\begin{equation}
\sigma_{n_{d}} = \left[-\frac{\partial \mu(n_d=1/2)}{\partial n_d}\right]^{-1/2}.
\label{finalwidth}
\end{equation}
It remains to compute $\partial \mu/\partial n_d$ for $n_d=1/2$. Taking the derivative
of the saddle point equations~(\ref{saddles2}), we can simplify them by inserting
$\mu=\nu=0$ at $n_d=1/2$, and we find:
\begin{eqnarray}
- I_2 \frac{\partial \nu}{\partial n_d} + \frac{1}{2} I_1 \frac{\partial \mu}{\partial n_d} & = & 0,\\
- I_1 \frac{\partial \nu}{\partial n_d} + \frac{1}{2} I_0 \frac{\partial \mu}{\partial n_d} & = & -2,
\end{eqnarray}
with $I_n = \int^{2\sqrt{N}\sigma}_{0} \text{d}E \, 
\frac{u(E) E^n}{(\lambda E^{2} + \rho)^2}$.
Clearly we have
the equality $\lambda I_2 + \rho I_0=N$ using the saddle point 
equation~(\ref{saddles2}) for $n_d=1/2$.
We obtain finally:
\begin{equation}
\label{derivative}
\frac{\partial \mu}{\partial n_d} = \frac{4 I_2}{(I_1)^2-I_0I_2},
\end{equation}
where the remaining integral are exactly computed:
\begin{eqnarray}
\label{I0}
I_0 &=& \frac{1}{32\lambda^2\sigma^2 \epsilon^{3/2}\sqrt{1+\epsilon}},\\
\label{I1}
I_1 &=& \frac{\sqrt{N}}{8\pi\lambda^2\sigma^3}\left[\frac{1}{\epsilon}-
\frac{1}{\sqrt{1+\epsilon}}\mr{arctanh}\left(\frac{1}{\sqrt{1+\epsilon}}\right)\right],\\
\label{I2}
I_2 &=& [N-(N/2-\lambda V^2)I_0]/\lambda.
\end{eqnarray}
The saddle point solution $\lambda = N/(2V^2)-1/(2\sigma^2)$ and
$\epsilon = (1/2-\lambda V^2/N)/(4\lambda\sigma^4)$ are both finite when $N\to\infty$,
since $V\propto \sqrt{N}$.
Inserting Eqs.~(\ref{I0})-(\ref{I2}) into Eq.~(\ref{derivative}) provides an analytical
expression for the width $\sigma_{n_{d}}$ of the charge distribution $P(n_d)$, given 
by Eq.~(\ref{finalwidth}). We find that this lengthy expression is very well approximated
in most of the range $V<\sqrt{2N}\sigma$, 
by the simple formula $\sigma_{n_{d}} = \sqrt{2/\pi-1/2} \sqrt{N}\sigma/V$, that
is asymptotically exact for $V\to \sqrt{2N}\sigma$.
The analytical form for the width $\sigma_{n_{d}}$ is compared in Fig.~\ref{fig:largeN} 
to the width extracted from the sampling method of Sec.~\ref{sec:numerics}, for several 
data as the one shown in the lower middle panel of Fig.~\ref{fig:numerics}, and
excellent agreement is found. In the weak coupling regime where $V\simeq \sigma/\sqrt{N}$,
deviations from the numerical data start to be seen, as expected since the large $N$ limit is 
only valid provided $V\propto \sqrt{N}$.

\section{Conclusion}
\label{sec:conclusion}

Recently, Bogomolny has considered a random matrix model modified by a local static potential~\cite{Bogomolny_2017},
corresponding to a rank-1 perturbation of the RMT. In this work, we have solved the problem of a random matrix
coupled to a dynamical quantum impurity (resonant level), which corresponds to a rank 2 perturbation
of the RMT.
In contrast with previous literature, we have considered typical observables of a fermionic system,
by investigating the electron charge distribution of the quantum impurity. We have first attacked
the problem using numerics (exact diagonalization of the impurity model and disorder averaging), 
and we found rich physical regimes.
When the impurity is weakly coupled to the band, a bimodal charge distribution with universal
(-3/2) power law was found. Although disorder is symmetrically centered around energy zero, 
the impurity has a tendency to polarize near zero or full occupation, due to its high sensitivity 
to charge noise in the weak coupling regime. 
At intermediate coupling, the impurity becomes diluted in the bath due to efficient hybridization, 
and a broad Gaussian charge distribution emerges. At strong coupling, the impurity forms a bound states 
with a single fermionic orbital of the bath, and the Gaussian distribution rapidly narrows down. 
All these observations are fully accounted for by analytical insights, obtained from the limit of either 
small or large number of sites in the bath. A general probability distribution functional was also exactly 
derived for the random matrix quantum impurity model. 

Perspectives of this work are manifold. First, there are still some open questions on the
simple spinless random quantum impurity model investigated in our manuscript, for instance regarding 
the statistics of the highest energy level in the band. Indeed, at small hybridization $V$, the
last level is unperturbed and follows the Tracy Widom distribution. However, at large hybridization
$V$, the level exits the band and displays Gaussian statistics. Describing the transition between
the two regimes would be interesting. Also at a theoretical level, it would
be more realistic to extend the calculations to the case of spinful fermions, and investigate
spin and charge distribution functions of the impurity level, depending also on the type of disorder
in the bath (random charge potentials, or random local magnetic fields). Since interactions are crucial 
to induce spin/charge separation (such as the phenomena of Kondo effect and Coulomb blockade), this study 
could be performed by a combination of analytical methods, as proposed here, and many-body
numerics, such as based on natural orbitals~\cite{Debertolis_2022}. 
Although our model based on Random Matrix Theory is over-simplified, we found by numerical analysis of a more 
realistic disordered tight-binding Hamiltonian that all our results were robust (not shown).
Thus, on a more experimental side, our study could be designed in artificial nanostructures, using a
small spin-polarized quantum dot coupled to a large chaotic billiard. Biasing electrostatic gates to modify 
the billiard shape, and a quantum point contact to measure the charge on the quantum dot impurity would allow 
to build the statistics necessary to confirm the predictions made in our paper. Careful calibration 
of the potential acting on the quantum dot would however be required to avoid trivial polarization effects.

\emph{Acknowledgements}.
We thank Weitao Chen, Izak Snyman, and Denis Ullmo for useful discussions.
MD was supported by the Deutsche Forschungsgemeinschaft through the cluster of excellence ML4Q (EXC 2004,
project-id 390534769) and by the Deutsche Forschungsgemeinschaft through CRC 1639 NuMeriQS (project-id 511713970).

\appendix

\section{Derivation of the probability distribution functions}
\label{AppA}

In this section, we derive the main result Eq.~(\ref{final_dist}) of Sec.~\ref{sec:PDF}, namely
the full probability distribution function of the eigenenergies and eigenstates of the
random matrix quantum impurity model.

\subsection{Definition of the eigenvalue problem}

We recall that our quantum impurity model is given by the $(N+1)\times(N+1)$ hermitian random matrix 
$H_{ij}$ in Eq.~(\ref{model_rmt}), which is a rank 2 perturbation to a standard $N\times N$ GOE 
random matrix $G_{ij}$. The latter is
extended here to size $(N+1)\times(N+1)$ in order to accomodate the decoupled impurity. 
We introduce $\ra{i}{\alpha}^{ }$ (resp. $\za{i}{\alpha}$) as the $i\text{th}$ component 
of the $\alpha\mathrm{th}$ eigenvector of $G_{ij}$ (resp. $H_{ij}$) with energy $e^{}_{\alpha}$ 
(resp. $E^{}_{\alpha}$), which obey: 
\begin{eqnarray}
\label{initeigen}
\som{j=0}{N} H^{}_{ij} \za{j}{\alpha} &=& \; E^{}_{\alpha}\za{i}{\alpha},\\
\som{j=0}{N} G^{}_{ij} \ra{j}{\alpha} &=& \; e^{}_{\alpha}\ra{i}{\alpha}, 
\,\text{with} \, e^{}_{0} = \epsilon^{}_{d},
\end{eqnarray}
where the decoupled impurity wave function
$\ra{0}{\alpha} = \di{0\alpha}$ with energy $e^{}_0 = \epsilon_d$ was added 
on the component $i=0$ of the $G_{ij}$ matrix.
The orthonormality of the eigenvectors $\phi$ reads:
\begin{equation}
\begin{cases}
\som{i=0}{N} \ra{i}{\alpha}^{ }\ra{i}{\beta} = \di{\alpha\beta}, \\
\som{\alpha=0}{N} \ra{i}{\alpha}^{ }\ra{j}{\alpha} = \di{ij},
\end{cases}
\end{equation}
and similarly for $\psi$.
An orthogonal transformation matrix $C_{\alpha\beta}=C^{-1}_{\beta\alpha}$ changes 
between the \textit{new} basis $\psi$ and the \textit{old} basis $\phi$ as:
\begin{eqnarray}
\label{changebasis1}
\za{j}{\alpha} &=& \som{\beta}{N}C^{}_{\alpha\beta} \, \ra{j}{\beta}, \\
\label{changebasis2}
\ra{j}{\alpha} &=& \som{\beta}{N}C^{-1}_{\alpha\beta} \, \za{j}{\beta}.
\end{eqnarray}
The problem is now completely defined, and we seek now explicit relations between the old set of
variables and the new ones in order to express the problem in the eigenbasis of Hamiltonian $H_{ij}$.
The probability distribution function being defined through integrals over the old variables, the determinant 
of the Jacobian associated to the change of basis needs to be calculated. This derivation is now presented step 
by step, in the spirit of Ref.~\cite{Bogomolny_2017}, which developed a similar yet simpler calculation
for a perturbation of rank 1 to a random matrix.

\subsection{Change of basis}

\subsubsection{From old eigenvectors \texorpdfstring{$ \{\ra{1}{\alpha}\}$}{TEXT} to new eigenvalues
\texorpdfstring{$\{ E^{}_{\alpha}\}$}{TEXT}}

The desired change of variables from old to new variables, $\left(\{e^{}_{\alpha}\},\{\phi^{}_{\alpha}\}\right) 
\rightarrow \left(\{E^{}_{\alpha}\},\{\psi^{}_{\alpha}\} \right)$, cannot be done in one shot, and hence will be
done in two steps. In this first section, eigenvectors of the decoupled impurity problem will be expressed through
the eigenvalues of the perturbed problem, this choice being relevant given the form of the
equations. We first start with the full eigenvalue problem~(\ref{initeigen}) in the new basis, 
and express it with the old basis:
\begin{eqnarray}
&&
\nonumber
\som{j=0}{N}\left[G^{}_{ij} + V(\di{i0}\di{j1} + \di{i1}\di{j0}) \right]\za{j}{\alpha} = E^{}_{\alpha}\za{i}{\alpha}, \\
&&
\nonumber
\som{\beta=0}{N}C^{}_{\alpha\beta}\left[ E_{\alpha}\ra{i}{\beta} 
- \som{j=0}{N}G^{}_{ij}\ra{j}{\beta}\right] = 
V[\di{i1}\za{0}{\alpha}\\
\nonumber
&&\hspace{6.8cm} + \di{i0}\za{1}{\alpha} ], \\
&&
\som{\beta=0}{N}C^{}_{\alpha\beta}\left(E^{}_{\alpha}-e^{}_{\beta} \right)\ra{i}{\beta} 
= V[\di{i1}\za{0}{\alpha} + \di{i0}\za{1}{\alpha} ].
\label{eq:change}
\end{eqnarray}
Multiplying the last equation by $\phi_i(\alpha)$, summing over $i$, and using orthogonality, we get
a simple expression for the transformation matrix:
\begin{eqnarray}
&& C^{}_{\alpha\beta} = V\frac{\ra{1}{\beta}^{ }\za{0}{\alpha} + \ra{0}{\beta}^{ }\za{1}{\alpha}}{E^{}_{\alpha}-e^{}_{\beta}}.
\end{eqnarray}
This last equality is injected in the change of basis \eqref{changebasis1}:
\begin{eqnarray}
\nonumber
&&\za{0}{\alpha} = V\som{\beta=0}{N}\frac{ \ra{1}{\beta}^{}\ra{0}{\beta}\za{0}{\alpha} +
|\ra{0}{\beta}|^{2}\za{1}{\alpha}}
{E^{}_{\alpha}-e^{}_{\beta}}, \\
\nonumber
&&\za{1}{\alpha} = V\som{\beta=0}{N}\frac{\ra{0}{\beta}^{}\ra{1}{\beta}\za{1}{\alpha} +
|\ra{1}{\beta}|^{2}\za{0}{\alpha}}
{E^{}_{\alpha}-e^{}_{\beta}}. \\
\end{eqnarray}
Using the wave function of the decoupled impurity problem, namely $\ra{0}{\alpha}=\di{\alpha0}$
and $\ra{1}{0}=0$, we obtain:
\begin{eqnarray}
\label{WF}
&&\za{0}{\alpha} = \frac{V}{E^{}_{\alpha}-\epsilon^{}_{d}} \; \za{1}{\alpha}, \\
&&\za{1}{\alpha} = \som{\beta=0}{N}\frac{V}{E^{}_{\alpha}-e^{}_{\beta}} |\ra{1}{\beta}|^{2}\za{0}{\alpha},
\end{eqnarray}
which are consistent provided that:
\begin{equation}
\left(E^{}_{\alpha}-\epsilon^{}_{d}\right) = \som{\beta=0}{N}\frac{V^{2}}{E^{}_{\alpha}-e^{}_{\beta}}
|\ra{1}{\beta}|^{2}.
\label{eigval0}
\end{equation}
Since $\ra{1}{0}=0$ is fixed, there are $N$ old wave function variables on one side $(|\ra{1}{1}|^{2},\hdots,|\ra{1}{N}|^{2})$, 
that are expressed in terms of the $N+1$ new energy variables $(E^{}_{0},\hdots,E^{}_{N})$, thus one of the variables $E^{}_{\alpha}$ 
is not independent. For the moment, this extra variable is not explicitely determined, and is written
$E_{0}=f(\{E^{}_{\alpha\geq1}\},\{\za{i\in\{1,0\}}{\alpha}\})$. The change of variables
$\ra{1}{\alpha>0} \rightarrow E^{}_{\alpha}$ is done first, so that $e^{}_{\alpha}$ and
$\za{i\in\{1,0\}}{\alpha}$ are supposed to be fixed for now. The system to be
solved Eq.~(\ref{eigval0}) can be rewritten in matrix form $\hat{A}\;\vec{b} = \vec{\Lambda}$,
where we note the vector 
$^{t}\vec{\Lambda}=\big((E_{1}^{}-\epsilon_{d}),\hdots,(E^{}_{N}-\epsilon_{d})\big)$, the
vector $^{t}\vec{b} =V^{2}\big( |\ra{1}{1}|^{2}, \hdots, |\ra{1}{N}|^{2}\big)$,
and $\hat{A}$ is a Cauchy matrix defined as $A^{}_{ij} = \frac{1}{E^{}_{i}-e^{}_{j}}$. 
Cramer's rule is used to determine the solutions $|\ra{1}{\beta}|^{2}$ through determinants of
matrices:
\begin{equation}
\label{Cramers}
V^{2}|\ra{1}{\beta}|^{2} = \frac{\det(\hat{A}_{\beta})}{\det(\hat{A})},
\end{equation}
with $\hat{A}^{}_{\beta}$ a $N\times N$ matrix defined by the subsitution of the column $\beta$ of $\hat{A}$ by $\vec{\Lambda}$:
\begin{equation}
\hat{A}^{}_{\beta} =
\begin{pmatrix}
\frac{1}{E^{}_{1}-e^{}_{1}} & \hdots & E^{}_{1}-\epsilon^{}_{d} & \hdots & \frac{1}{E^{}_{1}-e^{}_{N}} \\
\vdots & \ddots & \vdots & \ddots & \vdots \\
\vdots & \ddots & \vdots & \ddots & \vdots \\
\frac{1}{E^{}_{N}-e^{}_{1}} & \hdots & E^{}_{N}-\epsilon^{}_{d} & \hdots & \frac{1}{E^{}_{N}-e^{}_{N}} \\
\end{pmatrix}.
\end{equation}
The determinant of $\hat{A}_{\beta}$ is calculated with the Laplace expansion of the $\beta\mathrm{th}$ column:
\begin{equation}
\det(\hat{A}^{}_{\beta}) = \som{\alpha\geq1}{N}
(-1)^{\alpha+\beta}(E^{}_{\alpha}-\epsilon^{}_{d})\det(\hat{A}_{\beta\alpha}^{}),
\end{equation}
where $\hat{A}_{\beta\alpha}^{}$ is a matrix of rank $N-2$, defined by $\hat{A}$ without its row $\alpha$ and its column $\beta$. 
The determinant of $\hat{A}$ is also computed:
\begin{equation}
\det(\hat{A}) = \det
\begin{pmatrix}
\frac{1}{E^{}_{1}-e^{}_{1}} & \hdots & \frac{1}{E^{}_{1}-e^{}_{N}} \\
\vdots & \ddots & \vdots \\
\frac{1}{E^{}_{N}-e^{}_{1}} & \hdots & \frac{1}{E^{}_{N}-e^{}_{N}} \\
\end{pmatrix}.
\end{equation}
The column $\beta$ is subtracted to each column $j\neq\beta$, which leaves the determinant unchanged:
\begin{eqnarray}
\nonumber
\frac{1}{E_{i}^{}-e^{}_{j}} - \frac{1}{E_{i}^{}-e^{}_{\beta}} 
&=& \frac{E_{i}^{}-e^{}_{\beta} - E_{i}^{} + e^{}_{j}}{(E_{i}^{}-e^{}_{j})(E_{i}^{}-e^{}_{\beta})} \\ 
&=& \frac{(e^{}_{j} - e^{}_{\beta})}{(E_{i}^{}-e^{}_{\beta})(E_{i}^{}-e^{}_{j})}.
\end{eqnarray}
Each column is factorized by the term $(e^{}_{j} - e^{}_{\beta})$, and each row by the
factor$\frac{1}{E_{i}^{}-e^{}_{\beta}}$, so that:
\begin{eqnarray}
\nonumber
\det(A) &=& \pro{\gamma\geq1}{} \frac{1}{E_{\gamma}^{}-e^{}_{\beta}} \pro{\substack{\gamma\geq1, \\
\gamma\neq\beta}}{}(e^{}_{\gamma}-e^{}_{\beta})\\ 
&& \times \det
\begin{pmatrix}
\label{cauchy1}
\frac{1}{E^{}_{1}-e^{}_{1}} & \hdots & 1 & \hdots & \frac{1}{E^{}_{1}-e^{}_{N}} \\
\vdots & \ddots & \vdots & \ddots & \vdots \\
\vdots & \ddots & \vdots & \ddots & \vdots \\
\frac{1}{E^{}_{N}-e^{}_{1}} & \hdots & 1 & \hdots & \frac{1}{E^{}_{N}-e^{}_{N}} \\
\end{pmatrix}
\!\!.\,\,
\end{eqnarray}
Now the row $\alpha$ has to be eliminated, and it is subtracted to each row $i$ except $i=\alpha$:
\begin{eqnarray}
\nonumber
\frac{1}{E_{i}^{}-e^{}_{j}} - \frac{1}{E_{\alpha}^{}-e^{}_{j}} &=&
\frac{E_{\alpha}^{}-e^{}_{j} - E_{i}^{} + e^{}_{j}}{(E_{\alpha}^{}-e^{}_{j})(E_{i}^{}-e^{}_{j})}\\
&& = \frac{(E^{}_{\alpha} - E^{}_{i})}{(E_{\alpha}^{}-e^{}_{j})(E_{i}^{}-e^{}_{j})},\;\;\;\;
\end{eqnarray}
and as before each column (except column $\beta$) is factorized by
$\frac{1}{E^{}_{\alpha}-e^{}_{j}}$, and each row (except row $\alpha$) 
by $(E^{}_{\alpha} - E^{}_{i})$:
\begin{eqnarray}
\nonumber
\det(\hat{A}) &=& \frac{ \pro{\substack{\gamma\geq1, \\ \gamma\neq\beta}}{} (e^{}_{\gamma}-e^{}_{\beta}) 
\pro{\substack{\gamma\geq1, \\ \gamma\neq\alpha}}{} (E^{}_{\alpha} - E^{}_{\gamma})}
{\pro{\gamma\geq1}{}(E_{\gamma}^{}-e^{}_{\beta}) \pro{\substack{\gamma\geq1, \\ 
\gamma\neq\beta}}{} (E_{\alpha}^{}-e^{}_{\gamma})} \\ 
&& \times \det
\begin{pmatrix}
\frac{1}{E^{}_{1}-e^{}_{1}} & \hdots & 0 & \hdots & \frac{1}{E^{}_{1}-e^{}_{N}} \\
\vdots & \ddots & \vdots & \ddots & \vdots \\
1 & \ddots & 1 & \ddots & 1 \\
\vdots & \ddots & \vdots & \ddots & \vdots \\
\frac{1}{E^{}_{N}-e^{}_{1}} & \hdots & 0 & \hdots & \frac{1}{E^{}_{N}-e^{}_{N}} \\
\end{pmatrix}
\!\!,\,\,\\
\nonumber
&=& \frac{ \pro{\substack{\gamma\geq1, \\ \gamma\neq\beta}}{} (e^{}_{\gamma}-e^{}_{\beta}) \pro{\substack{\gamma\geq1, \\ \gamma\neq\alpha}}{} 
(E^{}_{\alpha} - E^{}_{\gamma})}
{\pro{\gamma\geq1}{}(E_{\gamma}^{}-e^{}_{\beta}) \pro{\substack{\gamma\geq1}}{} (E_{\alpha}^{}-e^{}_{\gamma})} 
(E_{\alpha}^{}-e^{}_{\beta}) (-1)^{\alpha+\beta} \\
&& \times \det(\hat{A}^{}_{\beta\alpha}),
\label{detAba}
\end{eqnarray}
using again Laplace expansion.
Putting everything back into \eqref{Cramers}, this equation is now fully expressed in terms
of the new energies $E_\alpha$:
\begin{equation}
\label{16}
V^2|\ra{1}{\beta}|^{2} = \frac{\pro{\gamma\geq1}{}(E_{\gamma}^{}-e^{}_{\beta}) }
{\pro{\substack{\gamma\geq1, \\ \gamma\neq\beta}}{} (e^{}_{\gamma}-e^{}_{\beta})}
\! \som{\alpha\geq1}{} \!
\frac{(E^{}_{\alpha} - \epsilon^{}_{d})}{(E_{\alpha}^{}-e^{}_{\beta})}
\frac{\pro{\substack{\gamma\geq1}}{} (E_{\alpha}^{}-e^{}_{\gamma}) }{\pro{\substack{\gamma\geq1, \\ \gamma\neq\alpha}}{} 
(E^{}_{\alpha} - E^{}_{\gamma})}.
\end{equation}
The sum (denoted $\mathbb{S}$) in the right hand side of Eq.~\eqref{16} can be expanded:
\begin{eqnarray}
\nonumber
\mathbb{S} &\equiv& \som{\alpha\geq1}{} \;
\frac{(E^{}_{\alpha} - \epsilon^{}_{d})}{(E_{\alpha}^{}-e^{}_{\beta})}
\frac{\pro{\substack{\gamma\geq1}}{} (E_{\alpha}^{}-e^{}_{\gamma}) }{\pro{\substack{\gamma\geq1, \\ \gamma\neq\alpha}}{} 
(E^{}_{\alpha} - E^{}_{\gamma})} \\ 
&=&
\som{\alpha\geq1}{} \;
\frac{(E^{}_{\alpha} - \epsilon^{}_{d})\pro{\substack{\gamma\geq1, \\ \gamma\neq\beta}}{} 
(E_{\alpha}^{}-e^{}_{\gamma}) }{\pro{\substack{\gamma\geq1, \\ \gamma\neq\alpha}}{} (E^{}_{\alpha} - E^{}_{\gamma})} \\
&=& \som{\alpha\geq1}{} \frac{E^{N}_{\alpha} - \left[\som{\gamma\geq1}{N}e^{}_{\gamma} - e^{}_{\beta} + \epsilon^{}_{d}\right]E^{N-1}_{\alpha} 
+ P(E^{}_{\alpha})}{\pro{\substack{\gamma\geq1, \\ \gamma\neq\alpha}}{} (E^{}_{\alpha} - E^{}_{\gamma})}.
\nonumber
\end{eqnarray}
Simple arguments from complex analysis and contour integration can be used to compute such sums~\cite{KnuthArt} as:
\begin{equation}
\label{simpsum}
\som{j=1}{N}\frac{x_{j}^{r}}{\pro{\substack{i=1, \\ i\neq j}}{N}(x^{}_{j}-x^{}_{i})} =
\begin{cases}
0, \quad \text{for} \quad 0\leq r < N-1, \\
1, \quad \text{for} \quad r = N-1, \\
\som{k=1}{N}x^{}_{k}, \quad \text{for} \quad r = N,
\end{cases}
\end{equation}
The expression for $\mathbb{S}$ is split into the sum of three terms: the first two are respectively proportional 
to $E^{N}_{\alpha}$ and $E^{N-1}_{\alpha}$, and the last one corresponds to a polynomial $P(E^{}_{\alpha})$ of
degree strictly smaller than $N-1$. Using the simplifications of Eq.~\eqref{simpsum}, Eq.~\eqref{16} becomes: 
\begin{eqnarray}
\nonumber
V^2 |\ra{1}{\beta}|^{2} &=& 
\frac{\pro{\gamma\geq1}{}(E_{\gamma}^{}-e^{}_{\beta}) }{\pro{\substack{\gamma\geq1, \\ \gamma\neq\beta}}{} 
(e^{}_{\gamma}-e^{}_{\beta})}
\; \som{\alpha\geq1}{} \left(E^{}_{\alpha} - e^{}_{\alpha} + e^{}_{\beta} - \epsilon^{}_{d} \right) \\
&=& \frac{\pro{\gamma\geq1}{}(E_{\gamma}^{}-e^{}_{\beta}) } 
{\pro{\substack{\gamma\geq1, \\
\gamma\neq\beta}}{} (e^{}_{\gamma}-e^{}_{\beta})} \; \left(e^{}_{\beta} - E^{}_{0}\right),
\label{defra}
\end{eqnarray}
where, to obtain the last line, we used $e_0=\epsilon_d$, as well as the trace property:
\begin{equation}
\som{i=0}{N} e^{}_{\alpha} = \text{Tr}(G) 
= \text{Tr}(H) = \som{i=0}{N} E^{}_{\alpha}.
\label{TraceH}
\end{equation}
The eigenvectors of the decoupled impurity problem are now totally determined in terms of the old
and new eigenvalues. Denoting the square amplitudes 
$r^{}_{\beta} = |\ra{1}{\beta}|^{2}$,
the determinant of the Jacobian for this change of variables is now computed for
$\alpha,\beta\geq1$:
\begin{equation}
\big[\hat{J}^{}_{\phi\rightarrow E}\big]_{\alpha, \beta} = \frac{\partial \, r^{}_{\beta}}{\partial E^{}_{\alpha}}.
\end{equation}
Computing the derivative, we get:
\begin{equation}
\det \left( \hat{J}^{}_{\phi\rightarrow E} \right) = \det \left(\frac{r^{}_{\beta}}{E^{}_{\alpha}-e^{}_{\beta}} +
\frac{\pro{\gamma\geq1}{} (E^{}_{\gamma}-e^{}_{\beta})}{V^{2}\pro{\substack{\gamma\geq1, \\
\gamma\neq\beta}}{} (e^{}_{\gamma}-e^{}_{\beta})} \right),
\end{equation}
using the fact that $\partial E_0 / \partial E_\alpha=-1$ for $\alpha\geq1$ from the
trace formula~(\ref{TraceH}).
The second term is identified from Eq.~\eqref{defra}:
\begin{equation}
\frac{\pro{\gamma\geq1}{} (E^{}_{\gamma}-e^{}_{\beta})}{V^{2}\pro{\substack{\gamma\geq1, \\ \gamma\neq\beta}}{} (e^{}_{\gamma}-e^{}_{\beta})} = \frac{-r^{}_{\beta}}{E^{}_{0} -
e^{}_{\beta}},
\end{equation}
so that the determinant is straightforwardly calculated:
\begin{eqnarray}
\label{detJ1}
\det \left( \hat{J}^{}_{\phi\rightarrow E} \right) 
&=& \det \left(\frac{r^{}_{\beta}}{E^{}_{\alpha}-e^{}_{\beta}} - \frac{r^{}_{\beta}}{E^{}_{0} - e^{}_{\beta}} \right)\\ 
\nonumber
&=& \det \left(\frac{r^{}_{\beta}(E^{}_{0}-E^{}_{\alpha})}{(E^{}_{0} - e^{}_{\beta})(E^{}_{\alpha}-e^{}_{\beta})} \right) \\
\nonumber
&=& \pro{\gamma\geq1}{}\left(\frac{E^{}_{0} - E^{}_{\gamma}}{E^{}_{0} - e^{}_{\gamma}}\right) 
\pro{\gamma\geq1}{}r^{}_{\gamma}\det\left(\frac{1}{E^{}_{\alpha}-e^{}_{\beta}} \right) \\ 
\nonumber
&=& \pro{\gamma\geq1}{}\left( \frac{E^{}_{0} - E^{}_{\gamma}}{E^{}_{0} - e^{}_{\gamma}}\right)
\frac{(-1)^{N}}{V^{2N}}\frac{\pro{\delta,\gamma\geq1}{}(E_{\gamma}^{}-e^{}_{\delta}) }
{\pro{\substack{\delta,\gamma\geq1, \\ \gamma\neq\delta}}{}
(e^{}_{\gamma}-e^{}_{\delta})}\\
\nonumber 
&&\times \pro{\delta\geq1}{}(E^{}_{0}-e^{}_{\delta})\frac{\pro{1\leq\gamma<\delta}{} (E^{}_{\delta}-E^{}_{\gamma})
(e^{}_{\gamma}-e^{}_{\delta})}{\pro{\gamma,\delta\geq1}{}(E^{}_{\gamma}-e^{}_{\delta})} \\
\nonumber
&=& \frac{1}{V^{2N}} \pro{1\leq\gamma<\delta}{} \frac{(E^{}_{\delta}-E^{}_{\gamma})}{(e^{}_{\gamma}-e^{}_{\delta})}
\pro{\gamma\geq1}{}\left(E^{}_{\gamma} - E^{}_{0} \right),
\end{eqnarray}
where we used standard property of the Cauchy determinant.
We thus have all the needed elements to replace the old eigenvectors of the decoupled impurity 
problem by the new eigenvalues of the perturbed problem in the probability distribution function. 
In order to close the problem in the complete set of new variables, we will now replace the
old eigenvalues of the decoupled problem by the perturbed new eigenvectors, using the same strategy.

\subsubsection{From old eigenvalues \texorpdfstring{$\{ e^{}_{\alpha}\}$}{TEXT} 
to new eigenvectors \texorpdfstring{$\{\za{0}{\alpha}\}$}{TEXT}}

The orthogonality of the matrix $\hat{C}$ is used to directly compute its inverse:
\begin{equation}
C^{-1}_{\alpha\beta} = C^{}_{\beta\alpha} = V\frac{\ra{1}{\alpha}^{ }\za{0}{\beta} 
+ \ra{0}{\alpha}^{ }\za{1}{\beta}}{E^{}_{\beta}-e^{}_{\alpha}}.
\end{equation}
As before, the previous equation is reinjected in \eqref{changebasis2}:
\begin{eqnarray}
\nonumber
\ra{0}{\alpha} &=& V \som{\beta=0}{N} \frac{|\za{0}{\beta}|^{2}\ra{1}{\alpha} + \za{0}{\beta}
\za{1}{\alpha} \ra{0}{\alpha}} {E^{}_{\beta} -e^{}_{\alpha}}, \\
\nonumber
\ra{1}{\beta} &=& V \som{\beta=0}{N} \frac{\za{0}{\beta} \za{1}{\beta} \ra{1}{\alpha} +
|\za{1}{\beta}|^{2}\ra{0}{\alpha}} 
{E^{}_{\beta}-e^{}_{\alpha}}.\\
\end{eqnarray}
Using $\ra{0}{\alpha}=\di{\alpha0}$, $\ra{1}{0}=0$, 
and the previous relation $\za{0}{\alpha} = \frac{V}{E^{}_{\alpha}-\epsilon^{}_{d}} \; \za{1}{\alpha}$
from Eq.~\eqref{eigval0}, we deduce the set of equations:
\begin{eqnarray}
\som{\beta=0}{N} |\za{0}{\beta}|^{2} &=& 1, \label{32a}\\
\som{\beta=0}{N} \frac{E^{}_{\beta}-\epsilon^{}_{d}}{E^{}_{\beta}-e^{}_{\alpha}} |\za{0}{\beta}|^2 &=& 1, \label{32b}\\
\som{\beta=0}{N} \frac{E^{}_{\beta} -\epsilon^{}_{d}}{V} |\za{0}{\beta}|^{2} &=& 0, \label{32c}\\
\som{\beta=0}{N} \frac{V}{E^{}_{\beta} -e^{}_{\alpha}}
|\za{0}{\beta}|^{2} &=& 0,\label{32d}
\end{eqnarray}
As before, there are
$N+1$ new variables $\za{0}{\beta}$ and only $N$ old variables $e^{}_{\alpha}$. Conditions on $\za{0}{0}$ to
reduce the number of variables are determined by Eq.~\eqref{32a}, that fixes the normalization of
the wave function. The condition fixing $E^{}_{0}$ was not set before, and is now given by
Eq.~\eqref{32c}. The two constraints are therefore:
\begin{eqnarray}
\label{constraintPsi0}
&& \som{\beta\geq0}{} |\za{0}{\beta}|^{2} = 1, \\
&&\som{\beta=0}{N} E^{}_{\beta} |\za{0}{\beta}|^{2} = \epsilon^{}_{d}.
\label{constraintE0}
\end{eqnarray}
where we have used Eq.~(\ref{constraintPsi0}) to simplify Eq.~(\ref{constraintE0}).
In order to express $|\za{0}{\beta}|^{2}$ for all $\beta\geq0$, we note that
Eq.~\eqref{32a} is the same equation as Eq.~\eqref{32b} for $\alpha=0$, such we can write
a single equation for both cases:
\begin{equation}
\som{\beta\geq 0}{} \frac{E^{}_{\beta}-\epsilon^{}_{d}}{E^{}_{\beta} -e^{}_{\alpha}} |\za{0}{\beta}|^{2} = 1, \quad \forall \alpha \geq 0.
\end{equation}
We thus found a similar equation to Eq.~\eqref{eigval0}, although it involves here a square $(N+1)\times (N+1)$ matrix
including the variables $z^{}_{0}$ and $E^{}_{0}$. The system to solve is:
\begin{equation}
\label{sysza}
\Big(E^{}_{\beta}-\epsilon^{}_{d} \Big)z^{}_{\beta} = \frac{\det(\hat{A}^{}_{\beta})}{\det(\hat{A})}, \quad \text{with} \quad z^{}_{\beta} = |\za{0}{\beta}|^{2}.
\end{equation}
In this case, the matrix $\hat{A}^{}_{\beta}$ is defined by:
\begin{equation}
\hat{A}^{}_{\beta} =
\begin{pmatrix}
\frac{1}{E^{}_{0}-e^{}_{0}} & \hdots & 1 & \hdots & \frac{1}{E^{}_{0}-e^{}_{N}} \\
\vdots & \ddots & \vdots & \ddots & \vdots \\
\vdots & \ddots & \vdots & \ddots & \vdots \\
\frac{1}{E^{}_{N}-e^{}_{0}} & \hdots & 1 & \hdots & \frac{1}{E^{}_{N}-e^{}_{N}} \\
\end{pmatrix}.
\end{equation}
As for Eq.~\eqref{cauchy1}, it is found that:
\begin{equation}
\det(\hat{A}) = \frac{\pro{\substack{\gamma\geq 0, \\ \gamma\neq\beta}}{}(E^{}_{\beta}-E^{}_{\gamma})}{\pro{\gamma\geq0}{} (E_{\beta}^{}-e^{}_{\gamma})} \det(\hat{A}^{}_{\beta}),
\end{equation}
and injecting into Eq.~\eqref{sysza}, we get:
\begin{eqnarray}
\label{defz}
 z^{}_{\beta} &=& \frac{\pro{\gamma\geq 1}{} (E_{\beta}^{}-e^{}_{\gamma})}{\pro{\substack{\gamma\geq0, \\ \gamma\neq\beta}}{}
(E^{}_{\beta}-E^{}_{\gamma})}.
\end{eqnarray}
We thus obtained the change of variable from the $N$ old variables $e^{}_{\alpha}$ 
to the $N+1$ new variables $z^{}_{\alpha}$,
and the constraint Eq.~\eqref{constraintE0} is enforced to supress the extra degree of freedom $z^{}_0$.
As before, the determinant of the Jacobian matrix associated to this change of variables needs to be 
determined. The derivative from Eq.~(\ref{defz}) is readily computed:
\begin{equation}
\left( \frac{\partial z^{}_{\beta}}{\partial e^{}_{\alpha}}\right)^{}_{\alpha,\beta\geq1} = \frac{-z^{}_{\beta}}{E^{}_{\beta}-e^{}_{\alpha}},
\end{equation}
so that by inserting Eq.~(\ref{defz}), its determinant reads:
\begin{equation}
\det(\hat{J}^{}_{e\rightarrow \Psi}) = \det \left( \frac{\partial z^{}_{\beta}}{\partial
e^{}_{\alpha}} \right)^{\!\!-1}
\!\!\!\!\!\!=(-1)^N \pro{\gamma}{}\frac{1}{z^{}_{\gamma}}
\det\left|\frac{1}{E_\beta-e_\alpha}\right|^{-1}\!\!\!\!\!\!,
\end{equation}
The Cauchy determinant above can be simplified by inserting Eq.~(\ref{defz}):
\begin{eqnarray}
\nonumber
\det(\hat{J}^{}_{e\rightarrow \Psi}) &=& (-1)^{N} \!\!
\left(\! \pro{\beta\geq1}{}z^{}_{\beta} \frac{ \pro{1\leq\delta<\gamma}{} (E^{}_{\delta}-E^{}_{\gamma}) 
(e^{}_{\gamma}-e^{}_{\delta})} {\pro{\delta,\gamma}{} (E^{}_{\delta}-e^{}_{\gamma})}\right)^{-1} \\
\nonumber
&=& (-1)^{N} \!\!\pro{1\leq\delta<\gamma}{}
\frac{(E^{}_{\delta}-E^{}_{\gamma})}{(e^{}_{\gamma}-e^{}_{\delta})} \pro{\delta\geq1}{}
(E^{}_{\delta}-E^{}_{0}) \\
\label{detJ2}
\end{eqnarray}
We arrived at the desired result, namely a complete change of variables from the old
variables (energies and wave functions) to the new ones. 
However, some terms of the wanted probability distribution function have still to be replaced 
using this change of variables. First, the original RMT distribution~(\ref{distrib0}) 
involves the sum of the squared eigenvalues of the decoupled impurity problem. 
Second, it also involves a delta function enforcing the normalization of the wave function,
namely the sum of every $r^{}_{\beta}$. Simple results are now found for these two terms.

\subsubsection{Sum of squared eigenvalues} \label{sec_som_e}
We aim here to express, in terms of the new variables, the sum of the old eigenvalues squared (decoupled 
impurity problem) which appears in the exponential in Eq.~\eqref{distrib0}.
This term is easily derived using the fact that the trace of the square matrix $\hat{G}$ does not to 
depend on the basis choice. We define $W_{ij} = V(\di{i1}\di{j0} + \di{i0}\di{j1})$, so that:
\begin{eqnarray}
\nonumber
\som{\alpha=0}{N}e^{2}_{\alpha} &=& \text{Tr}(G^{2}) = \text{Tr}(H-W)^2 = \text{Tr}(H^2-2WH+W^2)\\
&=& \som{\alpha=0}{N} E^{2}_{\alpha} + 2V^{2} - 4V\som{\alpha=0}{N} E^{}_{\alpha}
\psi^{}_{0}(\alpha) \psi^{}_{1}(\alpha),
\end{eqnarray}
where we have inserted $H_{ij}=\som{\alpha=0}{N} E^{}_{\alpha} \psi^{}_{i}(\alpha)
\psi^{}_{k}(\alpha)$.
From Eq.~\eqref{WF}, the sum is expressed with the new variables:
\begin{equation}
\label{som_esq}
\som{\alpha=0}{N}e^{2}_{\alpha} 
= \som{\alpha=0}{N} E^{2}_{\alpha} - 4\som{\alpha=0}{N} 
E^{}_{\alpha}(E^{}_{\alpha}-\epsilon^{}_{d}) z^{}_{\alpha} + 2V^{2},
\end{equation}
in terms of the amplitudes $z^{}_\alpha = |\psi_0(\alpha)|^2$.

\subsubsection{Sum of squared eigenvectors}

The sum of the amplitudes of the all old eigenvectors appears 
in Eq.~\eqref{distrib0} inside a Dirac delta function, to ensure the normalization of 
$r^{}_\beta=|\ra{1}{\beta}|^2$. 
This sum must also be expressed as a function of the new variables.
Using the property that $r^{}_{0} = 0$ (decoupled impurity) and Eq.~(\ref{defra}), 
the wanted term reads:
\begin{eqnarray}
\nonumber
\som{\beta=0}{N} r^{}_{\beta} &=& \som{\beta=1}{N} r^{}_{\beta} = \som{\beta=1}{N}
\frac{1}{V^{2}}\frac{\pro{\gamma\geq1}{}(E_{\gamma}^{}-e^{}_{\beta})}{\pro{\substack{\gamma\geq1, \\ \gamma\neq\beta}}{} 
(e^{}_{\gamma}-e^{}_{\beta})} \left(e^{}_{\beta} - E^{}_{0} \right) \\
\nonumber
&=& -\frac{1}{V^{2}} \som{\beta=1}{N} \frac{\pro{\gamma\geq0}{}(E_{\gamma}^{}-e^{}_{\beta})}{\pro{\substack{\gamma\geq1, \\ \gamma\neq\beta}}{}
(e^{}_{\gamma}-e^{}_{\beta})} \\
\nonumber
&=&-\frac{1}{V^{2}} \som{\beta=1}{N} 
\frac{(-1)^{N}}{(-1)^{N-1}\pro{\substack{\gamma\geq1, \\ \gamma\neq\beta}}{}
(e^{}_{\beta}-e^{}_{\gamma})} 
\Bigg[-e^{N+1}_{\beta}\\ 
\nonumber
&& + \som{\alpha=0}{N}E^{}_{\alpha}e^{N}_{\beta}
- \som{0\leq\alpha<\delta}{N}E^{}_{\alpha}E^{}_{\delta}e^{N-1}_{\beta} +
P(e^{}_{\beta}) \Bigg]
\end{eqnarray}
where $P(e^{}_{\beta})$ is a polynomial of order $N-2$, such that its sum vanishes
due to Eq.~\eqref{simpsum}. From this equation, we also get the simplification:
\begin{eqnarray}
\nonumber
\som{\beta=0}{N} r^{}_{\beta} 
&=& \frac{-1}{V^{2}} \som{\beta=1}{N} \frac{e^{N+1}_{\beta}}{\pro{\substack{\gamma\geq1, \\ \gamma\neq\beta}}{} 
(e^{}_{\beta}-e^{}_{\gamma})} + \frac{1}{V^{2}} 
\left(\som{\alpha=0}{N}E^{}_{\alpha}\right)\left(\som{\alpha=1}{N}e^{}_{\alpha}\right)\\
&& - \frac{1}{V^{2}}  \left(\som{0\leq\alpha<\delta}{N}E^{}_{\alpha}E^{}_{\delta}\right).
\label{almostthere}
\end{eqnarray}
However, the sum involving the power $e^{N+1}_{\beta}$ remains to be calculated.
For this purpose, the following complex valued function $f(z)$ is defined:
\begin{equation}
f(z) = \frac{z^{N+1}_{\pd}}{\pro{i=1}{N}(z-e^{}_{i})}.
\end{equation}
Using the residue theorem, this function can be integrated over a contour $\mathcal{C}$ enclosing
every simple pole $e^{}_{i}$:
\begin{equation}
\label{residue_fz}
\int\limits_{\mathcal{C}} \frac{\text{d}z}{2\pi i} \, f(z) \, \som{j=1}{N}\lim\limits_{z\rightarrow e^{}_{j}}(z-e^{}_{j})f(z) 
= \som{j=1}{N} \frac{e^{N+1}_{j} }
{\pro{\substack{i=1 \\ i\neq j}}{N} (e^{}_{i}-e^{}_{j}) },
\end{equation}
that is in fact the sum that needs to be computed. The function $f(z)$ is
analytic, so that a Laurent series can be defined on the contour $\mathcal{C}$ that will 
converge to $f(z)$:
\begin{equation}
f(z) = z \pro{i=1}{N}
\frac{1}{\left(1-\frac{e^{}_{i}}{z}\right)}=
 z \pro{i=1}{N} \left(\som{j=0}{+\infty} \left(\frac{e^{}_{i}}{z}\right)^{j}\right).
\end{equation}
The terms at first order in $\frac{1}{z}$ are developed
explicitely to compute the integral, which gives:
\begin{eqnarray}
\nonumber
\int\limits_{\mathcal{C}} \frac{\text{d}z}{2\pi i}\, f(z)  &=& \int\limits_{\mathcal{C}} 
\frac{\text{d}z}{2\pi i}\, z \pro{i=1}{N} 
\Bigg[ 1 + \frac{e^{}_{i}}{z} + \left(\frac{e^{}_{i}}{z}\right)^{2} 
+ o\left(\frac{1}{z^{2}}\right) \Bigg] \\
\nonumber
&=& \int\limits_{\mathcal{C}} \frac{\text{d}z}{2\pi i}\, \Bigg[ z + \left(\som{i=1}{N}e^{}_{i} \right)\\
\nonumber
&& + \left(\som{1\leq i<j}{N} e^{}_{i}e^{}_{j} + \som{i=1}{N}e^{2}_{i} \right)\frac{1}{z}
+ o\left(\frac{1}{z^{2}}\right) \Bigg] \\
&=& \left(\som{1\leq i<j}{N} e^{}_{i}e^{}_{j} + \som{i=1}{N}e^{2}_{i} \right),
\end{eqnarray}
where the residue theorem was used to obtain the final result (all other terms obviously cancel).
Comparing this last equation to Eq.~\eqref{residue_fz}, the desired result reads:
\begin{equation}
\som{j=1}{N} \frac{e^{N+1}_{j} }{\pro{\substack{i=1 \\ i\neq j}}{N} (e^{}_{i}-e^{}_{j}) } = \left(\som{1\leq i<j}{N} e^{}_{i}e^{}_{j} + \som{i=1}{N}e^{2}_{i} \right).
\end{equation}
From Eq.~(\ref{almostthere}), the sum of square amplitudes is thus:
\begin{eqnarray}
\nonumber
\som{\beta=0}{N} r^{}_{\beta} &=& \frac{-1}{V^{2}} \Bigg[ \som{i=1}{N}e^{2}_{i} + \som{1\leq i<j}{N} e^{}_{i}e^{}_{j} 
-\left(\som{\alpha=0}{N}E^{}_{\alpha}\right) \left( \som{\alpha=1}{N} e^{}_{\alpha}\right)\\
&& + \som{0\leq\alpha<\delta}{N} E^{}_{\alpha} E^{}_{\delta} \Bigg].
\end{eqnarray}
Further simplifications can be done using the rearrangement
$\som{1\leq i<j}{N} x^{}_{i} x^{}_{j} = \Big(\som{i=1}{N}x^{}_{i}\Big)^{2}/2
- \som{i=1}{N}x^{2}_{i}/2$
into the sums including $E^{}_{\alpha}E^{}_{\delta}$ and $e^{}_{i}e^{}_{j}$,
leading to the following:
\begin{eqnarray}
\nonumber
\som{\beta=0}{N} r^{}_{\beta} &=& \frac{-1}{V^{2}} \Bigg[ \frac{1}{2} \som{i=1}{N}e^{2}_{i} 
+ \frac{1}{2} \left(\som{i=1}{N}e^{}_{i}\right)^{2}\\
\nonumber
&-& \left(\som{\alpha=0}{N} E^{}_{\alpha}\right)
\left( \som{\alpha=1}{N} e^{}_{\alpha}\right)
+ \frac{1}{2}\left(\som{i=0}{N}E^{}_{i}\right)^{2}\!\! - \frac{1}{2}\som{i=0}{N}E^{2}_{i} \Bigg] \\
\nonumber
&=& \frac{-1}{2V^{2}} \left[ \left(\som{i=1}{N}e^{2}_{i} - \som{i=0}{N}E^{2}_{i} \right) 
+ \left(\som{i=1}{N}e^{}_{i} - \som{i=0}{N}E^{}_{i}\right)^{2} \right]\!.\\
\label{presque}
\end{eqnarray}
We inject Eq.~\eqref{som_esq} in the first term under parenthesis of (\ref{presque}), and
the second term under parenthesis cancels out due the trace formula~(\ref{TraceH}), so that
the normalization of the old wavefunction in terms of the new variables reads:
\begin{equation}
\label{som_rsq}
\som{\beta=0}{N} r^{}_{\beta} = \frac{2}{V^{2}} \som{\alpha=0}{N} E^{}_{\alpha}(E^{}_{\alpha}-\epsilon^{}_{d}) z^{}_{\alpha} - 1.
\end{equation}

\subsection{Full probability distribution}

Every part of the change of variables being now defined, the joint distribution of eigenvalues and
eigenvectors of the random impurity matrix $H$ can be derived:
\begin{equation}
P\Big(\{E^{}_{\alpha}\},\{z^{}_{\alpha}\}\Big) = |\det(\hat{J}^{}_{e\rightarrow \Psi}) 
\det(\hat{J}^{}_{\phi\rightarrow E})|
P\Big(\{e^{}_{\alpha}\},\{r^{}_{\alpha}\}\Big),
\end{equation}
with the original GOE random matrix distribution given in Eq.~(\ref{distrib0}).
From the Jacobian~(\ref{detJ1}) and the expression~(\ref{defra}) for the amplitudes,
the first change of variables reads:
\begin{eqnarray}
\nonumber
&&P\Big(\{e^{}_{\alpha}\},\{E^{}_{\alpha}\}\Big)
= |\det(\hat{J}^{}_{\phi\rightarrow E})| \, P\Big(\{e^{}_{\alpha}\},\{r^{}_{\alpha}\}\Big) \\
\nonumber
&& \;\;\;\; \propto 
\pro{1\leq\gamma<\delta}{} \frac{(E^{}_{\delta}-E^{}_{\gamma})}{(e^{}_{\gamma}-e^{}_{\delta})} 
 \pro{\gamma\geq1}{}\left(E^{}_{\gamma} - E^{}_{0}\right) 
\pro{1\leq\alpha<\beta}{N}(e^{}_{\alpha} - e^{}_{\beta}) \\ 
\nonumber
&& \;\;\;\; \times \frac{\pro{\substack{\gamma\geq1, \\ \gamma\neq\alpha}}{}
(e^{}_{\gamma}-e^{}_{\alpha})^{\frac{1}{2}}}{(e^{}_{\alpha} - E^{}_{0})^{\frac{1}{2}} \pro{\gamma\geq1}{}
(E_{\gamma}^{}-e^{}_{\alpha})^{\frac{1}{2}}}
\delta\!\left(\som{\gamma\geq1}{N}r^{}_{\gamma} - 1 \right) \! e^{-\frac{1}{4\sigma^2} \som{\gamma\geq1}{N} e_{\gamma}^{2}} \\
\nonumber
&& \;\;\;\; \propto \frac{\pro{1\leq\gamma<\delta}{} (E^{}_{\delta}-E^{}_{\gamma})(e^{}_{\delta} - e^{}_{\gamma})}
{\pro{\delta,\gamma\geq1}{}(E_{\gamma}^{}-e^{}_{\delta})^{\frac{1}{2}}}
\, \pro{\gamma\geq1}{} \frac{\left(E^{}_{\gamma} - E^{}_{0}\right)}{\left(e^{}_{\gamma} -
E^{}_{0}\right)^{\frac{1}{2}}}\\
&&\;\;\;\; \times \delta\left(\som{\gamma\geq1}{N}r^{}_{\gamma}-1\right) \! e^{-\frac{1}{4\sigma^2} \som{\gamma\geq1}{N} e_{\gamma}^{2}}.
\end{eqnarray}
Using Jacobian~(\ref{detJ2}) for the second change of variables:
\begin{widetext}
\begin{eqnarray}
\nonumber
P\Big(\{E^{}_{\alpha}\},\{z^{}_{\alpha}\}\Big)
&=& |\det(\hat{J}^{}_{e\rightarrow \Psi})| 
P\Big(\{e^{}_{\alpha}\},\{E^{}_{\alpha}\}\Big) \\
\nonumber
&\propto& \frac{\pro{1\leq\gamma<\delta}{} (E^{}_{\delta}-E^{}_{\gamma})(e^{}_{\delta} -
e^{}_{\gamma})}{\pro{\delta,\gamma\geq1}{}(E_{\gamma}^{}-e^{}_{\delta})^{\frac{1}{2}}} \,
\!\pro{\gamma\geq1}{} \frac{\left(E^{}_{\gamma} - E^{}_{0}\right)}{\left(e^{}_{\gamma} -
E^{}_{0}\right)^{\frac{1}{2}}} \delta\left(\som{\gamma\geq1}{N}r^{}_{\gamma}-1\right)\!\! 
\pro{1\leq\delta<\gamma}{}
\frac{(E^{}_{\delta}-E^{}_{\gamma})}{(e^{}_{\gamma}-e^{}_{\delta})}
\pro{\delta\geq1}{} (E^{}_{\delta}-E^{}_{0}) e^{-\frac{1}{4\sigma^2} \som{\gamma\geq1}{N} e_{\gamma}^{2}}\\
&\propto& \frac{\pro{1\leq\gamma<\delta}{} (E^{}_{\delta}-E^{}_{\gamma})^{2}}{\pro{\delta,\gamma\geq1}{}(E_{\gamma}^{}-e^{}_{\delta})^{\frac{1}{2}}} \, \pro{\gamma\geq1}{}
\frac{\left(E^{}_{\gamma} - E^{}_{0}\right)^{2}}{\left(e^{}_{\gamma} - E^{}_{0}\right)^{\frac{1}{2}}} 
\delta\left(\som{\gamma\geq1}{N}r^{}_{\gamma}-1\right)
e^{-\frac{1}{4\sigma^2} \som{\gamma\geq1}{N} e_{\gamma}^{2}},\\
\nonumber
&\propto& \pro{1\leq\gamma<\delta}{}
(E^{}_{\delta}-E^{}_{\gamma}) \, \pro{\gamma\geq1}{} \frac{\left(E^{}_{\gamma} -
E^{}_{0}\right)^{3/2}}{\left(e^{}_{\gamma} - E^{}_{0}\right)^{\frac{1}{2}}} e^{-\frac{1}{4\sigma^2} \som{\gamma=0}{N} E^{2}_{\gamma}} 
\pro{1\leq\alpha}{}\frac{1}{(z^{}_{\alpha} )^{\frac{1}{2}}} 
\delta\left(\som{\gamma=0}{N} E^{}_{\gamma}(E^{}_{\gamma}-\epsilon^{}_{d}) z^{}_{\gamma} - V^{2} \right)\\
\label{longeq}
&& \times
e^{\frac{1}{4\sigma^2}\left[4\som{\gamma=0}{N} E^{}_{\gamma}(E^{}_{\gamma}-\epsilon^{}_{d})
z^{}_{\gamma}-2 V^2 \right]}\\
&\propto& \pro{0\leq\gamma<\delta}{} (E^{}_{\delta}-E^{}_{\gamma}) \pro{0\leq\alpha}{}\frac{1}{(z^{}_{\alpha} )^{\frac{1}{2}}}
\delta\left(\som{\gamma=0}{N} E^{}_{\gamma}(E^{}_{\gamma}-\epsilon^{}_{d}) z^{}_{\gamma} - V^{2}
\right)e^{-\frac{1}{4\sigma^2} \som{\gamma=0}{N} E^{2}_{\gamma}},
\label{eqalmost}
\end{eqnarray}
\end{widetext}
by inserting Eq.~\eqref{som_rsq} in the delta function and Eq.~\eqref{som_esq} in the exponential
term. We also used $1/(z^{}_{0})^{\frac{1}{2}} = 
\pro{\gamma\geq 1}{} (E_{0}^{}-E^{}_{\gamma})^{1/2}/\pro{\gamma\geq1}{}(E^{}_{0}-e^{}_{\gamma})^\frac{1}{2}$
to obtain the last line, and the delta function was applied to cancel the second exponential term (up to a constant) 
in Eq.~(\ref{longeq}).
Finally, we impose the constraints on the square amplitudes $z^{}_0$ and energy $E_0$, since these
variables are not independent, as noted earlier.
These constraints, which have been derived previously in Eq.~(\ref{constraintPsi0})-(\ref{constraintE0}), 
simply read: $z^{}_{0} = 1 - \som{\gamma\geq1}{} z^{}_{\gamma}$ 
and $\som{\gamma \geq 0}{} E^{}_{\gamma} z^{}_{\gamma} = \epsilon^{}_{d}$. 
Finally gathering these constraints with Eq.~(\ref{eqalmost}), the full probability distribution of the 
random matrix quantum impurity problem can be written as in Eq.~(\ref{final_dist}).

\bibliographystyle{apsrev4-2}
%


\end{document}